\documentclass[12pt, preprint]{aastex}

\shorttitle{Photometry M31, M33}
\shortauthors{Massey et al.}
\slugcomment{Astronomical Journal, in press}

\begin{document}
\title{A Survey of Local Group Galaxies Currently Forming Stars: \\I. {\it UBVRI} Photometry of Stars in M31 and M33\altaffilmark{1}}

\author{Philip Massey,\altaffilmark{2, 3} 
K. A. G. Olsen,\altaffilmark{4}
Paul  W. Hodge,\altaffilmark{5}
Shay B. Strong,\altaffilmark{2,6}
George H. Jacoby,\altaffilmark{7}
Wayne Schlingman,\altaffilmark{8}
R. C. Smith\altaffilmark{4}}

\altaffiltext{1}{Based in part on observations made with the NASA/ESA Hubble Space Telescope,
obtained at the Space Telescope Science Institute, which is operated by the Association
of Universities for Research in Astronomy (AURA), Inc. under NASA contract NAS 5-26555.  These
observations are associated with program GO-9794.}
\altaffiltext{2}{Lowell Observatory, 1400 W. Mars Hill Rd., Flagstaff, Az 86001;
Phil.Massey@lowell.edu}
\altaffiltext{3} {Visiting Astronomer, Kitt Peak National Observatory, National Optical 
Astronomy Observatory (NOAO), which is operated by AURA, Inc., under cooperative agreement
with the National Science Foundation (NSF).}
\altaffiltext{4}{Cerro Tololo Inter-American Observatory (CTIO), NOAO, which is operated by AURA, Inc.,
under cooperative agreement with the NSF.}
\altaffiltext{5}{Department of Astronomy, University of Washington, Seattle WA 98195}
\altaffiltext{6}{Research Experiences for Undergraduates (REU) at CTIO, 2001. Present address: Department of Astronomy, RLM 16.318, University of Texas,  Austin, TX 78712-1083.}
\altaffiltext{7}{WIYN Observatory, P. O. 26732, Tucson, AZ 85726-6732.}
\altaffiltext{8}{Research Experiences for Undergraduates (REU) student at Lowell Obs., 2003.  Present
address: Steward Observatory, University of Arizona, 933 N. Cherry Ave., Tucson, AZ 85719.}

\begin{abstract}

We present {\it UBVRI} photometry obtained from Mosaic images of
M31 and M33 using the KPNO 4-m telescope.  We describe our data
reduction and automated photometry techniques in some detail, as
we will shortly perform a similar analysis of other Local Group
galaxies.  The present study covered 2.2 square degrees along the
major axis of M31, and 0.8 square degrees on M33, chosen so as to
include all of the regions currently active in forming massive
stars.   We calibrated our data using data obtained on the Lowell
1.1-m telescope, and this external method resulted in millimag
differences in the photometry of overlapping fields, providing some
assurance that our photometry is reliable. The final catalog contains
371,781 and 146,622 stars in M31 and M33, respectively, where every
star has a counterpart in (at least) the $B$, $V$, and $R$ passbands.
Our survey goes deep enough to achieve 1-2\% photometry at 21st
magnitude (corresponding to stars more massive than $20M_\odot$)
and achieves $<$10\% errors at $U\sim B \sim V \sim R \sim I \sim
23$rd mag.  Although our typical seeing was only modest (0.8-1.4",
median 1.0") by some standards, we find excellent correspondence
between our catalog sources and those we see in our {\it HST} ACS
data for OB48, a crowded region in M31.  We compare our final
photometry with those of others, and find good agreement with the
CCD catalog of M31 stars by Magnier et al., although our study
covers twice the area and goes about 2 mags deeper.  The photographic
studies of others fare less well, particularly at the faint end in
$V$, where accurate background subtraction is needed for good
photometry.  We provide cross references to the stars confirmed as
members by spectroscopy, and compare the location of these to the
complete set in color-magnitude diagrams.  While follow-up spectroscopy
is needed for many projects, we demonstrate the success of our
photometry in being able to distinguish M31/M33 members from
foreground Galactic stars.  Finally, we present the results of a
single night of spectroscopy on the WIYN 3.5-m telescope\footnote{
The WIYN Observatory is a joint facility of the University of
Wisconsin-Madison, Indiana University, Yale University, and NOAO.}
examining the brightest likely members of M31.  The spectra identify
34 newly confirmed members, including B-A supergiants, the earliest
O star known in M31, and two new Luminous Blue Variable candidates
whose spectra are similar to that of P~Cygni.

\end{abstract}


\section{Introduction}
\label{Sec-intro}

The galaxies of the Local Group serve as our astrophysical laboratories for studying the effects
that metallicity and other environmental factors have on star formation and massive star
evolution.  The advent of 4-m class telescopes and single-object spectrographs in the 1970s
heralded in an era of such studies in the Magellanic Clouds, where even the modest 
change in metallicity (a factor of 4 from the SMC to Milky Way) have resulted in some 
revealing
differences in the characteristics of the massive star populations, such 
as the distribution of spectral subtypes
of Wolf-Rayet stars (WRs)  and red supergiants (RSGs), as well as a 
factor of 100 difference
in the relative numbers of WRs and RSGs.  These differences are believed to be 
due to the effect that metallicity has on the mass-loss rates of massive stars, and the subsequent
large effect on stellar evolution.  (For a recent review, see Massey 2003.)  The observed ratios
can be used to test and 
``fine-tune" stellar evolution theory (see Meynet \& Maeder 2005), and so it
is important for such measurements to be extended to as low and high metallicities as possible.
With the introduction of multi-object
spectrographs on larger aperture telescopes (GMOS on Gemini, DEIMOS on Keck), it is now
possible to extend such studies to the more distant members of the Local Group, where the
galaxies forming stars span a range of 20 in metallicity (WLM to M31, see Table 1 of Massey 2003).

Such spectroscopic studies require the knowledge of an
appropriate sample of stars to observe.  We became
aware of the need for a comprehensive survey of the resolved stellar content of nearby galaxies
in support of our own research; we realized, however, that an additional strength of this study would
be the uses that other researchers could make of such data.   We took advantage of the 
NOAO ``Survey program" to use the Mosaic CCD cameras
on the KPNO and CTIO 4-m telescopes to image the Local Group galaxies currently actively
forming stars.  Our Local Group Galaxies Survey (LGGS) project imaged 
the nearby spirals M31 in ten fields,
and M33 in three fields, as well as the dwarf irregulars NGC 6822, WLM,
IC 10, the Phoenix Dwarf, Pegasus, Sextans A, and Sextans B, each in a single field.  
(The need to complete
M31 and M33 precluded the inclusion of IC 1613, which is located at a similar
right ascension, but we hope someday to correct the omission.)
The survey includes {\it UBVRI} data, as well as images through narrow-band (50-80\AA) filters
centered on H$\alpha$, [OIII] $\lambda 5007$, and
[S II] $\lambda 6713, 6731$.   Our goal was to obtain uniform large-area coverage of the
star-forming regions in these galaxies, with broad-band photometry good to 1-2\% 
for massive stars ($\ge 20M_\odot$).  The data would be taken under good, but not always
excellent, seeing conditions ($<1.0-1.2$ arcsec).    These data could be supplemented by
WFPC2/ACS images with {\it HST} or by AO for higher resolution studies of small
regions, but our survey would provide uniform coverage of
the entire galaxies.  The broad-band photometry would be used to characterize
the stellar population of massive stars in these galaxies.  By itself, it would separate
red supergiants (RSGs) from foreground dwarfs, and allow us to identify OB stars
for follow-up studies.  At intermediate colors, it would at least identify the
sample of stars that must be examined spectroscopically to identify F-G supergiants.
The narrow-band data
would be used to distinguish H$\alpha$ emission-line stars from planetary nebulae and supernovae
remnants.  

\begin{figure}
\epsscale{1.0}
\plotone{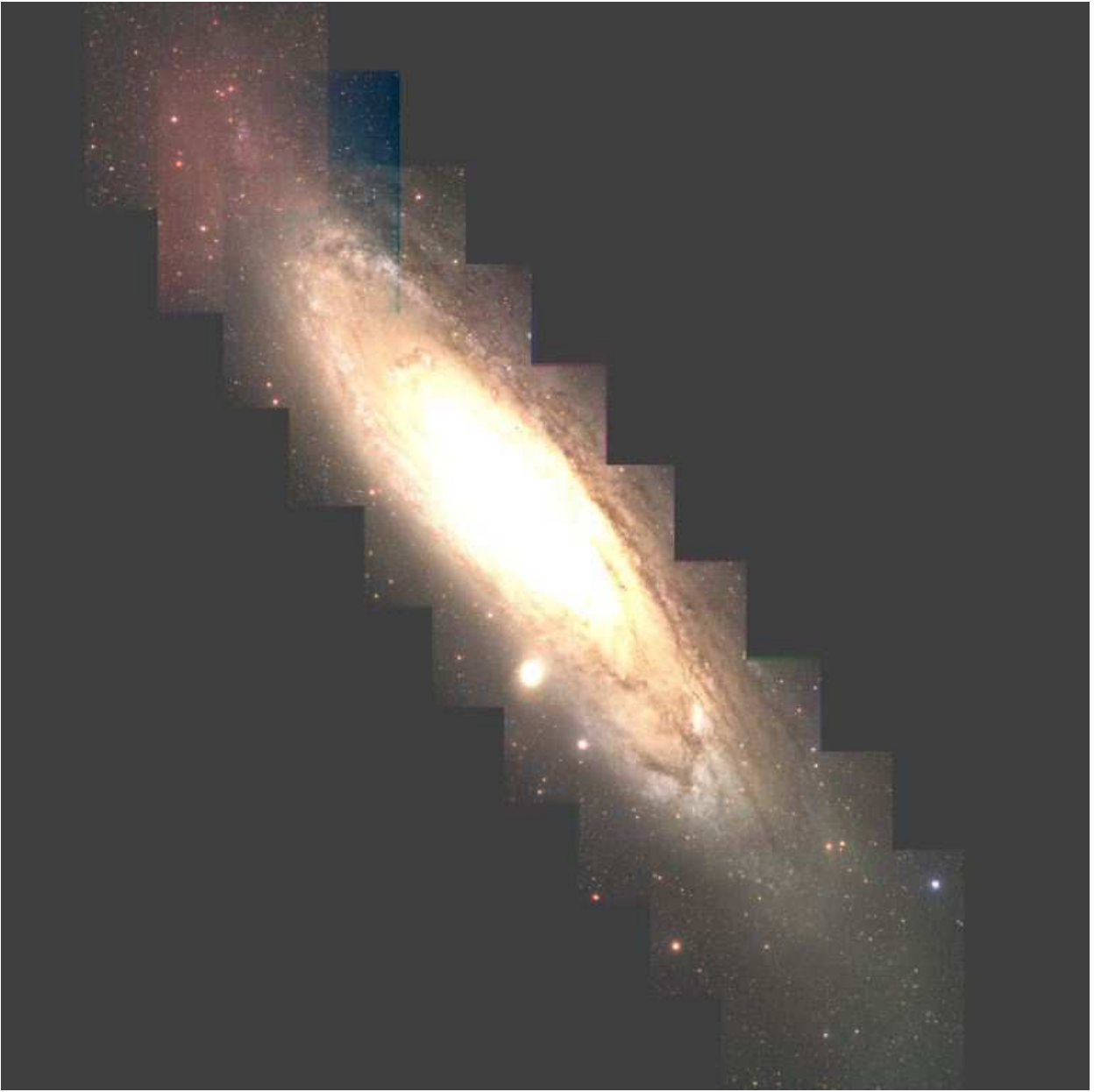}
\caption{\label{fig:M31} M31 Mosaic Fields.  We show here the ten fields
(numbered from upper left to lower right) that we used to cover the star-formation
regions of M31.  A slight (several ADU) problem with the flatness of the stacked
image of Field 2 (upper left)
results in the slightly different color; the problem did not affect
our photometry, which relies upon local sky determination.  Each Mosaic field
is roughly 36'x36' in size.  The coverage of M31 was designed to go beyond the
OB associations identified by van den Bergh (1964).
}
\end{figure}

\begin{figure}
\epsscale{0.8}
\plotone{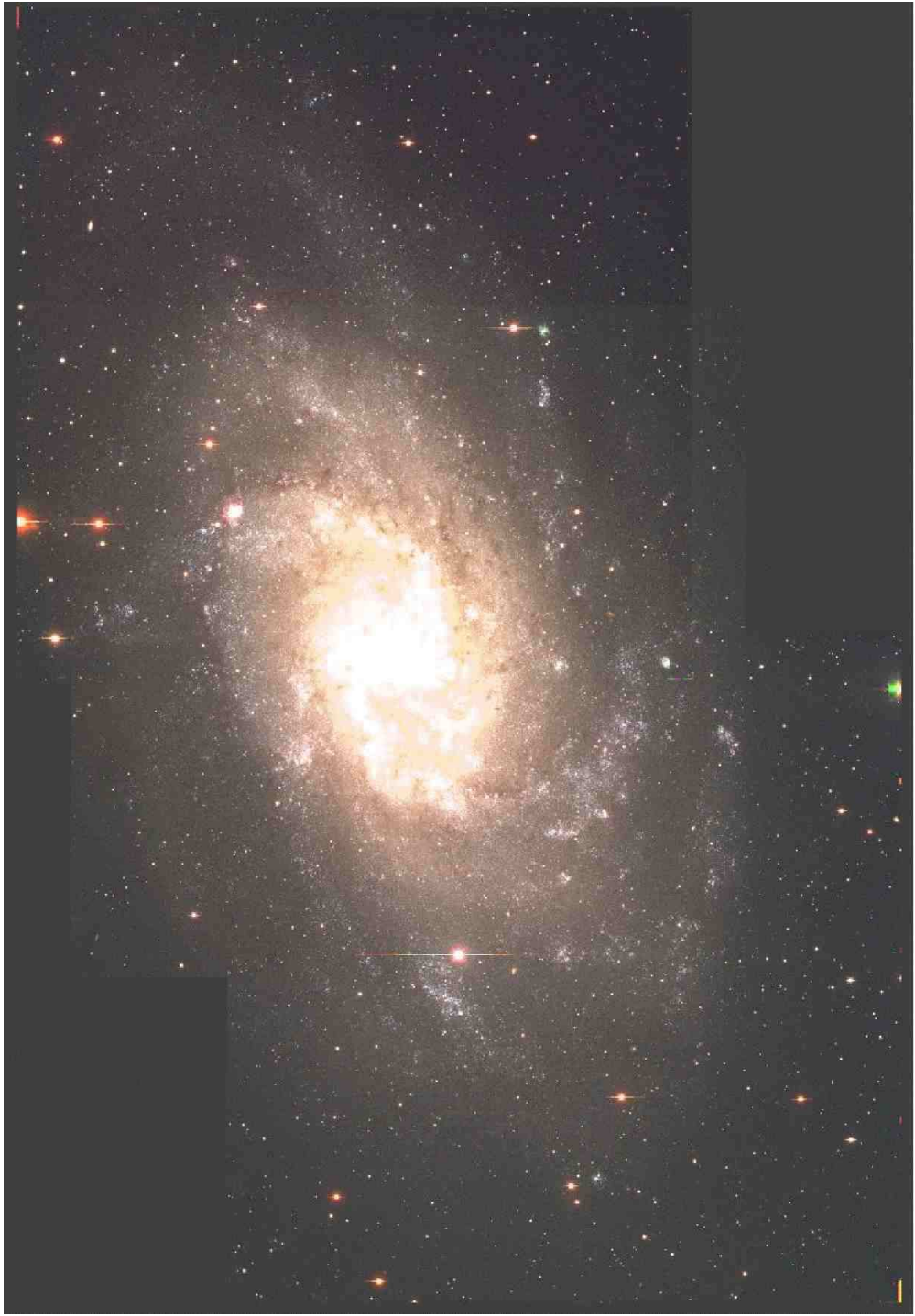}
\caption{\label{fig:M33} M33 Mosaic Fields.  We show here the three fields
(north, center, and south) that we used to cover the star-forming regions of
M33.  Our coverage extends well beyond the OB associations identified
by Humphreys \& Sandage (1980).
}
\end{figure}

The observing began in August 2000, and ran through September 2002, with a total of 16 nights of usable data 
obtained on the CTIO and KPNO 4-m telescopes.  Most of this time was spent on M31 and M33,
as these spirals occupied the largest amount of area on the sky (see Figs.~\ref{fig:M31} and \ref{fig:M33}).  
The complete set of images have been available since 2003 via the NOAO Science
Archive\footnote{http://archive.noao.edu/nsa/} and Lowell web sites\footnote{http://www.lowell.edu/users/massey/lgsurvey}.  
Here we present our {\it UBVRI} photometry of stars in M31 and M33.  Our survey covers 2.2 square
degrees in M31, and 0.8 square degrees in M33.  Subsequent
papers will describe the results of our emission-line filters, and our broad-band photometry of the dwarfs.
Our survey has already been used as part of two 
PhD projects (Williams 2003, Bonanos et al.\ 2006)
as well as other studies (Di Stefano et al.\ 2004, Massey 2006,
Humphreys et al.\ 2006).

In \S~\ref{Sec-data} we describe our data, and go into some detail into the reduction philosophy and technique, 
since the same methods have been applied to the complete data set.
In \S~\ref{Sec-results} we present the catalogs and compare our photometry to that of others.
Although spectroscopic follow-up is crucial for addressing many of our science drivers (the dependence
of the IMF and upper-mass cutoffs on metallicity, accurate physical H-R diagrams for comparison with
stellar evolutionary models), the photometric data alone can be used to good advantage, and we
illustrate this in \S~\ref{Sec-analysis} where we present color-magnitude diagrams 
(\S~\ref{Sec-CMD})
and illustrate the power of the photometry in identifying blue and red members (\S~\ref{Sec-BR}).
We test the findings using
a preliminary spectroscopic reconnaissance (\S~\ref{Sec-spectra}).  We summarize our results,
and describe our plans for future work in \S~\ref{Sec-future}.

\section{Observations and Data Reductions}
\label{Sec-data}

In Table~\ref{tab:journal} we list the field centers and observation dates for all of our M31 and M33 
Mosaic frames.  The data were taken with the Mosaic CCD camera at the prime focus of the
4-m Mayall telescope.  The camera consists of an array of 8 thinned 2048x4096 SITe CCDs with 15$\mu$m
pixels.  The scale is  0.261" pixel$^{-1}$ at the center, and decreases to 
0.245" pixel$^{-1}$ at the corners
due to pincushion distortion from the optical corrector (Jacoby et al.\ 1998).  To maintain good
image quality, an ADC is used during broad-band exposures.  A single exposure subtends
an area of the sky roughly 35' by 35'; however, there are gaps between the chips (3 gaps of 12" each in the
NS direction; 1 gap of 8" EW), and so the usual observing procedure is to obtain a set of 5 dithered
exposures with the telescope offset slightly (25-50") between each exposure.  The area covered by
a dither sequence is about 36' by 36'.

The basic reductions were performed with the ``mscred" package in 
IRAF\footnote{IRAF is distributed by NOAO, which is operated
by AURA, Inc., under cooperative agreement with the NSF.}.  The reductions are somewhat more
complicated than that of a normal (single) CCD; complete details can be found at the LGGS
web site. Complications include
the fact that there is appreciable cross-talk between pairs of chips that share the same electronics,
causing an electronic ``reflection" of a small fraction ($\leq$0.3\%)
of the signal of one chip to be added to that of the other.
This was most easily seen by the reflection of heavily saturated stars, but
if left uncorrected would have affected all of the data on half of the chips.
In addition, the corrector introduced a significant ($\leq$10\%) optical reflection
``ghost pupil" of the sky affecting
the central portions of the field.
Finally, the change of scale resulted in the need to rectify the images using stars with good
astrometric positions within the field.    

For each run we began by carefully determining the cross-talk terms using the 
nominal values and
revising these until we obtained good subtraction of saturated stars, as judged by eye.  Next, we constructed a bad-pixel
mask by dividing a long and short exposure of the dome flat.  This mask would be used to flag non-linear
pixels, which would then not be used in the photometry.   For each night we obtained bias frames,
dome flats, and sky flats.  Given the read-out time (130s) it was not practical to obtain twilight flats 
in each filter each night, but a good set was obtained on each run. Each set of biases and flats
were combined after cross-talk correction and over-scan removal.
A combination of dome flats and sky flats were needed to
construct an image of the ``ghost pupil" for each filter; this ghost was subtracted from the sky flats.
After these preliminaries were done, we proceeded as follows: (1) Cross-talk was removed using
our revised coefficients.  (2) The overscan was removed line-by-line for each chip, and each chip
``trimmed" to remove the overscan region.  (3)  A revised bad-pixel mask was constructed combining
the run-specific image and automatically determining saturated values and any bleed trails.
(4) The two-dimension
bias structure was removed by subtracting the average bias frame for the run. (5) The data were
flat-fielded using the (cleaned) average sky flats. (6) The filter-specific ghost image was fit to each
image, and interactively examined to determine the optimal scaling factor.  This correction was most
significant for the {\it U} and {\it I} exposures.  (7) An astrometric solution for each frame was 
performed using stars from the USNO-B1.0 catalog (Monet et al.\ 2003).  
The higher-order astrometric
distortion terms were left at their default values, but individual scales were determined for
each axis, as well as rotation. 
This solution was then used
to resample the data (using a time-consuming but robust sinc interpolation algorithm),
de-projecting the image to a constant scale (0.27" pixel$^{-1}$) with conventional astronomical
orientation (N up, E left) with a single tangent point for each galaxy.  This allows for simple
registration of adjacent fields, if desired.

For many users of the Mosaic camera, the ultimate goal of the basic reduction process
is the construction of a single ``stacked" image from the processed individual exposures;
this image is cosmetically clean, and can then be used for subsequent analysis.  We realized
at the beginning of the project, however, that this would not be adequate if we were to achieve our
goal of 1-2\% photometry through the broad-band filters.  A simple division of a {\it B}
and {\it V} dome flat suggests that there are highly significant differences in the color-terms between
the various chips.  The stacked image would contain star images that had been
combined from as many as 4 different chips as a result of dithering.
Therefore, we made the decision to treat each CCD separately in the photometric
analysis.  This did require 40x more work (8 CCDs $\times$ 5 ditherings) in general, but our sense
was that 
in the end we would have significantly better photometry.  The stacked images do suffice for the
analysis of our narrow-band (H$\alpha$, [OIII], [SII]) data,
which we will discuss in future papers.

This decision freed us from the wasteful task of observing broad-band standard stars with the 4-m.
Since the read-out time of the array is 130s, observing a single standard star offset to
each of 8 chips through 5 filters would require nearly 1.5 hours simply in read-out time.  One could not
hope to observe sufficient standards during a night for a precise photometric solution.  In addition, the
use of external calibration allowed us to make use of nights on the 4-m which were mostly clear, but
not completely photometric.  For the photometric calibration, 
we used the Hall 1.1-m telescope on Lowell Observatory's dark sky site on Anderson Mesa.
Data were obtained on 26 nights from 2000 November through 2003 February.  
The detector was a 2048 by 2048 SITe CCD with 24$\mu$m
pixels.  The chip was binned 2x2, with an image scale of 
1.14" pixel$^{-1}$, and a field of view of
19.4'$\times$19.4'. The seeing was 
typically 2-3".  For each M31 and M33 field, we obtained two exposures in each filter, with the
telescope offset by 500" north and 500" south of the Mosaic field centers 
(Table~\ref{tab:journal}).
This assured us that there would be overlap between the photometric
 frames and the area included on each of the Mosaic CCD frames.  Exposure times were 
 1200 s in {\it U}, 120 s in each of {\it B, V,} and {\it R}, and 300 s in {\it I}, chosen so there would
 be good overlap between stars with adequate counts on the calibrating frames and the brightest
 unsaturated stars on the Mosaic frames.  The allocation of observing time on the small telescope
 was sufficiently generous so that we could use only the best, photometric nights.  Typically
 each calibration field was bracketed by observations of a dozen or so of the best-calibrated (i.e., observed multiple times with errors less than 0.01~mag)
 Landolt (1992) standards.  This allowed us to determine extinction terms accurately; after most of
 the standard data were reduced, we fixed the values for the color-terms, and found optimal zero-points
 and extinction values for each night.  The average residuals for the
 standard solutions were 1-2\% for all filters.  As usual,
 we found that the {\it U} solutions required two different slopes; one for $U-B>0$, and one for
 $U-B<0$. (See Massey 2002 for discussions of difficulties with calibrating $U$-band photometry via
 CCDs and the standard UG2/UG1 + CuSO$_4$ {\it U}-like filters.)
 
 For the photometry, we developed scripts\footnote{Our full set of software, including IRAF scripts
 and FORTRAN code, can be downloaded from http://www.lowell.edu/users/massey/lgsurvey/splog2.html.  This software is offered ``as is",  with no implied warranty.} that separated each of the Mosaic dithered exposures
into the 8 individual chips, and characterized the exposure (median sky value, full-width-half-maximum 
 [FWHM]) and updated the headers (read-noise, saturation value, gain).   Our scripts relied
 upon the basic IRAF/DAOPHOT routines (Stetson et al.\ 1990), but performed the tasks automatically
 in order to deal with the huge data volumes.
 Stars 4$\sigma$ above the
 local background were found with the appropriate FWHM and image shapes, and aperture photometry
 was performed with a small (3.5 pixel radius) aperture.  This was done
 independently for each filter. Suitable PSF stars were automatically
 identified, and simultaneous PSF-fitting was performed over the frame using the ``allstar" algorithm.  
 Additional stars were added
 based upon residuals from subtracting the fitted PSFs from the original frame, and the simultaneous
 PSF-fitting was repeated.   Similar routines were run on the Lowell 1.1-m data, and isolated stars were
 matched between the data sets to determine photometric zero-points and color terms.  When all of
 the data were reduced, we then examined the results and fixed the color terms to the values given
 in Table~\ref{tab:colorterms}.  We then reran our calibration program to
 determine the best zero-points\footnote{Note that since the color-terms are
 not the same for each chip, the photometric
 zero-points will not be the same for each
 chip either, even though they were taken as a single image.  The reason is that the
flat-field lamps do not have zero color.  The error introduced by ignoring this
effect would be about 1-2\%.  Our calibration procedure explicitly found individual
zero-points for each chip on each image once the color terms were determined.}.
 
An examination of the variations of the color terms between
chips reveals that our concerns were well-founded.
Had we treated the chips as identical, we would have introduced a difference
of 0.06~mag in $U$ for a
lightly-reddened O star ($U-B\sim -1$) dithered between chips 4 and 5.  Similarly, a red supergiant
($B-V\sim 2$)  would have derived $B$ values that 
differed by 0.10 mag, depending upon whether the star
was observed on chip 3 or chip 5.  (The derived $B-V$ colors would have been less affected;
i.e., a difference of about 0.05~mag.) 
For projects requiring 5-10\% photometry, 
or narrow-band filters
(where color-terms are negligible) the use of stacked images would suffice,  but to be
able to achieve 1-2\% broad-band photometry (and not be limited by calibration issues) 
requires some extra work.

We averaged the calibrated photometry for each field, and then compared the differences in adjacent
fields, 
restricting the sample to only well-exposed stars (statistical uncertainty $<$1\%).  
The results are shown in Table~\ref{tab:difs}.
 Often the
median differences were only several millimag.   Note that this is a critical test of our
photometric accuracy, since each field was calibrated independently using external
data.
We were pleased to find evidence that we generally were table to reduce any
calibration issue to $<1$\%, even at $U$.

Before releasing our final catalogs, we made one additional step, that of removing
false detections along diffraction spikes, a problem that has plagued
other surveys as well (see, for example, Magnier et al.\ 1992).   Stars brighter
than (roughly) 13th~mag had noticeable diffraction spikes, oriented
 at 45$^\circ$ to the cardinal
directions.  For the brightest foreground stars (7th mag), these diffraction spikes
extended $\sim 200$ pixels from the star.  We found that around each bright
star there were a handful of false detections in our preliminary catalog.  For each
source near the coordinates of a bright star, we computed the ellipticity and position
angle of the object using the stacked $V$ image for convenience.  If the ellipticity
was high, and the position angle aligned towards  the bright star, we
eliminated the source from the catalog.  Checking the results visually, we seem to have
eliminated nearly all of such bogus detections, with little cost in terms of real
objects.  This affected only 0.1\% of the sources in the two catalogs, but removed
an annoyance.  

\section{Results}
\label{Sec-results}

\subsection{The Catalogs}
\label{Sec-cats}

The final catalog consists of the averaged photometry for each star; of course, many of these stars were observed
multiple times (i.e., on five ditherings and possibly on as many as three overlapping fields).  For a star to make it into the catalog,
it had to be detected in the $B$, $V$, and $R$ filters; thus there are stars without $U-B$ or $R-I$ measurements,
and we denote these null measurements with a magnitude of ``99.99".  The complete M31
and M33 catalogs are
available in machine-readable format via the on-line edition; 
in the printed versions of Tables~\ref{tab:M31} and \ref{tab:M33}
we present the first ten entries of each. The M31 catalog contains a total
of  371,781 stars, and the M33 catalog contains a total of 146,622 stars. The stars have been assigned designations
based on their celestial (J2000) positions; i.e., LGGS J004341.84+411112.0 refers to the star with coordinates
$\alpha_{\rm 2000}=00^{\rm h}43^{\rm m}41.\!^{\rm s}84$, $\delta_{\rm 2000}=+41^\circ11' 12.\!"0$ following IAU conventions. 
(This particular star is a very close analog of P~Cygni, and is discussed both
in Massey
2006 and below in \S~\ref{Sec-spectra}.)

How deep, and complete,  does our survey go? 
In terms of the photometric precision, we show the {\it median} errors as a function of magnitude
for the combined M31 and M33 catalogs
in Table~\ref{tab:errors}.  We see that the errors are $<$10\% through about 23rd magnitude\footnote{A few of the very brightest stars have slightly larger errors than some fainter stars.  This is due to the fact that the formal errors contain not only the photon and read-noise, but are also scaled by the reduced chi-squared value of fitting the PSF
to a particular star.  Since in general the PSF will be based upon the average of stars
slightly fainter than the brightest star on a frame, the errors of the brighter stars may
be higher than expected just based upon Poisson statistics.}

Our stated goal was to  achieve 1-2\% photometry for massive ($\geq 20 M_\odot$) stars.
Did we achieve this? Let us briefly consider the evolution of a $20 M_\odot$
star; for details see 
Massey (1998a). On the zero-age
main sequence the star would be identified as an O9.5~V star and have $M_V=-3.5$. 
The intrinsic colors of such a star will be $(U-B)_0\sim -1.1$, $(B-V)_0\sim -0.3$, $(V-R)_0 \sim -0.1$,
and $R-I\sim-0.2$ (Bessell et al.\  1998).  The observed
colors for such a star depend upon the reddening; let us assume that the star has an
$E(B-V)=0.25$, typical of several OB associations in M31 (Massey et al.\ 1986).  In that case,
we expect that such a star will have $U-B=-0.9$, $B-V=0.0$, $V-R=0.0$, and $R-I=0.0$.
Thus, at a distance modulus of 24.4 (M31, van den Bergh 2000), the star would have
$U=20.0$, $B=V=R=I=20.9$\footnote{M33 is another tenth of a magnitude further away
according to van den Bergh (2000), but the typical reddening of an OB star is less
(Massey et al.\ 1995).}.  The error in $R$ will be slightly larger than our goal 
(0.027 vs. 0.020 mag),
 but in all the other bands we will have achieved our goal; for early-type stars
it will be $UBV$ that is particularly useful as a temperature discriminant (Massey 1998a).
Some 8 million years later, near the end of its hydrogen-burning phase, the star would be
a B1~I, with $M_V=-5.3$, and nearly identical intrinsic colors, and easily within
our criteria.
Finally, as a He-burning object the star would be spectroscopically identified as a
RSG.  As an M0~I, the star
would have $M_V=-6.8$, with intrinsic colors of $(U-B)_0=2.5$,
$(B-V)_0=1.8$, $(V-R)_0=0.9$, and $(R-I)_0=0.8$, or $U-B=2.7$, $B-V=2.0$, $V-R=1.0$,
and $R-I=1.0$.  So, roughly $U=22.3$, $B=19.9$, $V=17.9$, $R=16.9$, and $I=15.9$.  The error
at $U$ ($\sigma \sim 0.04$)
is a little larger than our goal, but the others all give better than 1\% statistics.  We will 
see in \S~\ref{Sec-BR} the usefulness of good colors at this magnitude.

\begin{figure}
\epsscale{0.9}
\plotone{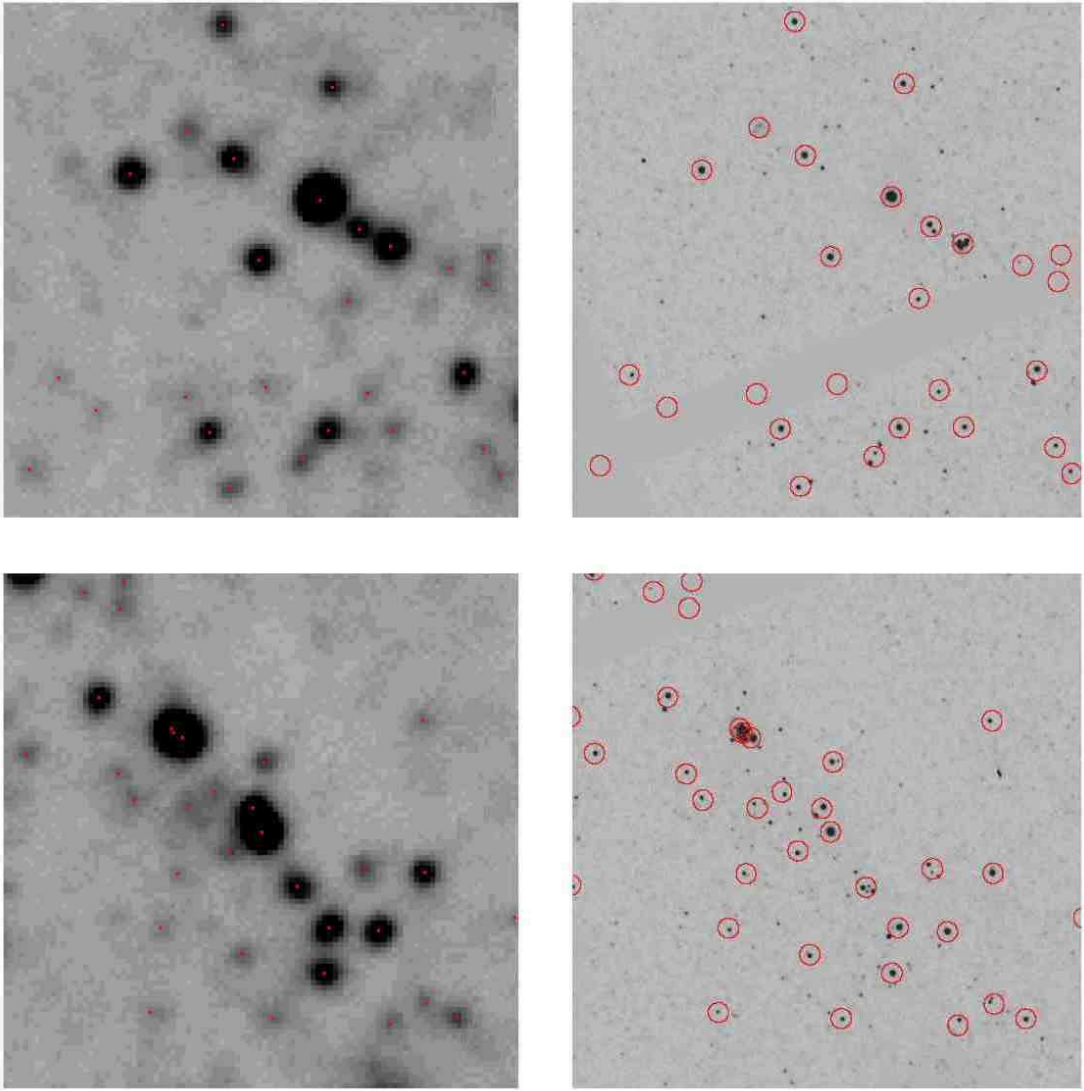}
\caption{\label{fig:OB48} Comparison of the LGGS Mosaic field of OB48 with
that taken by ACS (F555W filter).   
The images on the left show a small section of the {\it V} stacked image of
M31-F4 containing OB48, an association rich in massive stars (Massey et al.\ 1986).
The images on the right were obtained with the ACS wide-field camera (F555W filter)
on  {\it HST}, with a scale of 0.05" pixel$^{-1}$.  We have indicated the stars in
our catalog (Table~\ref{tab:M31}). We see that although there are stars that are
multiple at {\it HST} resolution, they are often (but not always!) marked detected as
multiple in our survey as well.  Each section is roughly 25" on a side.  The circles
on the ACS images have a diameter of 1.0".}

\end{figure}

However, a more critical test concerns how well we did in crowded regions.
Obviously there are stars in M31 and M33 that cannot be resolved from the
the Earth---this is true, after all, even for massive stars at 2~kpc distances.
But, we were of course curious how well we did in general.  In Fig.~\ref{fig:OB48}
we compare our $V$ stacked LGGS image of
 OB48, an association rich in massive
stars (Massey et al.\ 1986), with an ACS image shown to the same scale and
orientation.  We have indicated the stars in our M31 catalog.   We see that there
are a few cases where stars were multiple at {\it HST} resolution but detected as 
single objects in our survey.  But, generally our ground-based imaging
did very well.  We call
particular attention to the star at left of center in the upper pair.  That star is
OB48-444, one of the earliest known O star in M31 (Massey et al.\ 1995), an O8~I star.
Of course, it is possible to have unresolved multiple systems even at {\it HST}
resolution, but as Kudritzki (1998) has emphasized, such 
multiplicity usually reveals itself
by spectral peculiarities.

\subsection{Comparison with Others}

Photometry of galaxy-wide samples of the resolved stellar content of M31 and M33 have
mainly been carried out photographically; e.g., Berkhuijsen et al.\ (1988) for M31 and
Humphreys \& Sandage (1980) and Ivanov et al.\ (1993)
for M33.  Only the Magnier et al.\ (1992) survey of M31
has used CCDs in such a global study.  Others have imaged small areas of these galaxies
with CCDs from the ground
(for example, Massey et al.\ 1986, 1995; 
Hodge \& Lee 1988; Hodge et al.\ 1988; Wilson et al.\ 1990), or even smaller regions
using {\it HST} (Hunter et al.\ 1996a, 1996b; Magnier et al.\ 1997).

The CCD survey of Magnier et al.\ (1992) broke new ground by providing 
{\it BVRI} photometry of 361,281 stars in a 1 square degree area of M31\footnote{The number of stars in the complete 
Magnier et al.\ (1992) catalog is comparable to the number in
ours,
despite the fact our survey goes considerably deeper,
as we counted as valid only
stars that were matched in $B$, $V$, {\it and} $R$ in order to eliminate
spurious detections.  Stars in Magnier et al.\ (1992)
were usually detected only in a single filter.  
The number of stars that were
detected by Magnier et al.\ (1992) in $B$, $V$, and $R$ is 19,966, according
to their Table 2.}.
Indeed, this survey provided much of the inspiration for the present study.
We compare the properties of the two surveys in Table~\ref{tab:Magnier}.
Given our larger aperture telescope, and the improvement in the size of
CCD cameras in the past decade, it is not surprising that our survey goes 
about 2~mag deeper and covers about twice the area.  

\begin{figure}
\epsscale{1.0}
\plotone{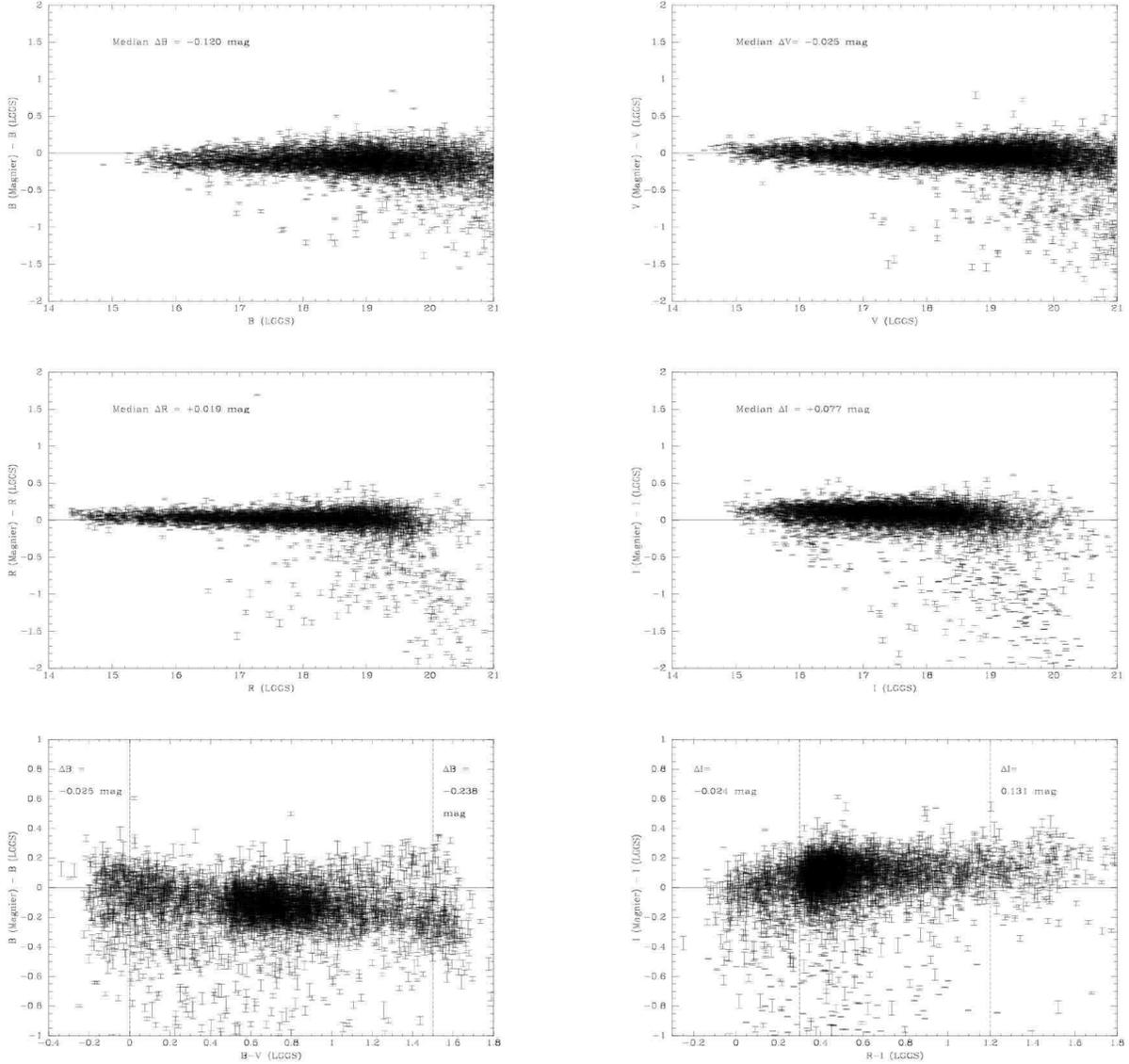}
\caption{\label{fig:magdif} 
Comparison between our photometry and that of Magnier et al.\ (1992) of M31.
In making this comparison we have restricted the sample to stars with photometric errors less
than 0.05 mag for each filter, and stars that are isolated (no significantly bright
neighbors within 10".)    The median differences are shown.  The differences
found for $B$ and $I$ seem to be color related, as shown in the bottom two panels.
}
\end{figure}

We compare our photometry to that of Magnier et al.\ (1992) in Fig.~\ref{fig:magdif}.
Since the image quality is so different (our worst seeing was their best), we restricted
the comparison to stars in our catalog that have no comparably bright companions 
($V_{\rm star} - V_{\rm comp} < 1$) within
10".   We have also restricted the comparison to the stars with the best photometry
(errors less than 0.05~mag in each filter), although nearly identical values
are obtained if we loosen or tighten this restriction.  We find median
differences (in the sense of Magnier et al.\ 1992 minus LGGS) of 
$-0.120$ in $B$ (5,191 stars),
$-0.025$ in $V$ (7,214 stars), +0.019 in $R$ (4,129 stars) and +0.077 in $I$ (5,387 stars).  
The differences
at $V$ and $R$ are small and as expected; the modest offset in $B$ and $I$ 
appear to be real.    As shown in the bottom two panels of Fig.~\ref{fig:magdif} the
reason for the differences at $B$ and $I$ appear to be color related: for the bluest
stars, our results and Magnier et al.\ (1992) are in good accord, while for the
reddest stars the differences are significantly larger. 
Stars with $B-V<0$ show a median difference of $-0.025$~mag, while
stars with $B-V>1.5$ show a median difference of $-0.238$~mag.  Similarly there
is a strong correlation of the $I$ differences with color, with the bluest stars
($R-I<0.3$) showing good agreement (median difference $-0.024$), while the
reddest stars ($R-I>1.2$) show a large difference (+0.131).
We are of course biased towards believing these color problems are
inherent to Magnier et al.\ (1992) and not our own data, but of course only 
independent observation can answer that.  We do note that we were careful to
include a full range of colors of standards in obtaining our secondary calibration,
while Magnier et al.\ (1992) relied upon published M31 photometry for their
calibration.
At least one of these sources,  Massey et al.\ (1986),
was well calibrated for only the bluest stars, and that may help explain some of
these differences.

\begin{figure}
\epsscale{1.0}
\plotone{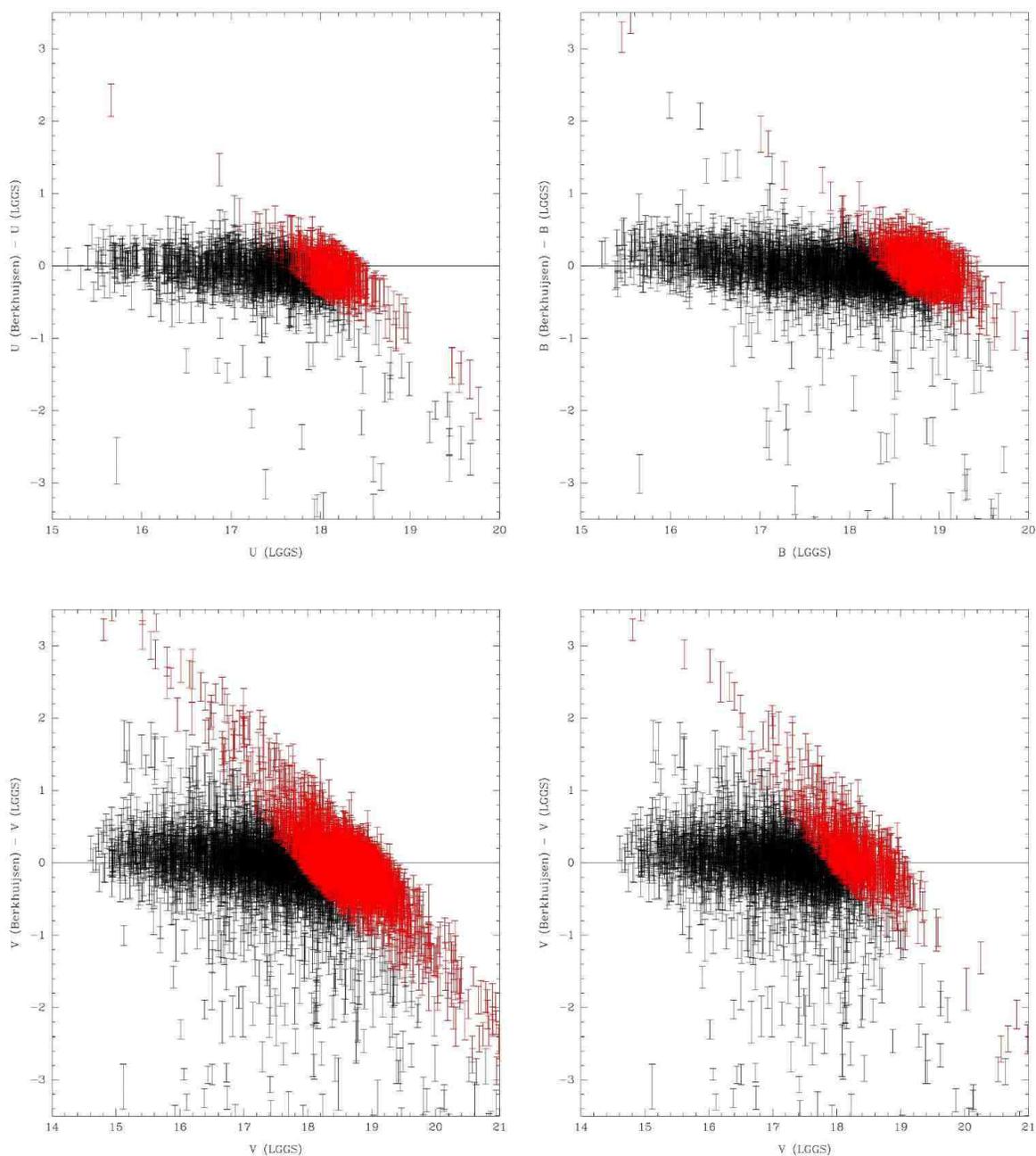}
\caption{\label{fig:berk}
Comparison between our photometry and that of Berkhuijsen et al.\ (1988) of M31.
The fainter stars in the Berkhuijsen et al.\ (1988) catalog
are marked in red (Berkhuijsen et al.\, $U>17.8$, $B>18.5$, $V>18.0$).
The $U$ and $B$ plots show relatively good agreement, but a large
systematic effect is present for the fainter stars in $V$,  amounting to several
magnitudes.  In the plot at bottom right, we make the comparison in $V$
photometry only to those stars that appear in the $B$ comparison.  Although
fewer data are present, the same trend is evident.  Since the same trend is not
evident in the $B$ plot, the problem cannot be due to mistaken matches.
}
\end{figure}

We also were curious to compare our results to the M31 photometry catalog
of Berkhuijsen et al.\ (1988),
based upon reductions of photographic plates.  Massey (2006) noted that there
appeared to be a significant magnitude-dependent difference between the {\it V}
magnitudes of Magnier et al.\ (1992) and those of Berkhuijsen et al.\ (1988), at least
in a small region around the star AF And.  The problem is complicated by the fact
that the coordinates in Berkhuijsen et al.\ (1988) are also known to contain 
large systematic errors, as noted by Magnier et al.\ (1992).  In comparing their
coordinates to ours, we find that we need to correct the 
Berkhuijsen et al.\ (1988) coordinates by $-0.1^s$ and $+2.5"$ to 
bring the averages into accord with ours,
and that in addition there were problems at the $\pm$5" level.  The median differences
in the photometry are in reasonable agreement:
$-0.093$ in $U$, $-0.046$ in $B$, and $-0.040$ in $V$.  However,
as we can see in Fig.~\ref{fig:berk} there is
a very strong effect with magnitude,
at $V$, with the faintest stars in Berkhuijsen et al.\ (1988)
shown in red. 
Such stars show systematic differences up to several magnitudes!
We were concerned that this sort of effect might be due to incorrect
matching of stars, given the problems in the Berkhuijsen et al.\ (1988) coordinates,
and so we also made the comparison to just those stars that had both $V$ {\it and} $B$.
This is shown in the bottom-right panel of Fig.~\ref{fig:berk}.  We see the same 
effect, although of course with fewer data.  Since the $B$ data do not show this
problem, we conclude it is not an issue with matching.
The problem we find here with the Berkhuijsen et al.\ (1988) photometry
is consistent with Massey (2006),
who found an $V=17.5$ (LGGS) star listed as 18.1 in Berkhuijsen et al.\ 
(1988), although the $B$ values agreed well.
 Not all of the Berkhuijsen et al.\ (1988) data are affected---there are plenty of fainter
stars that do agree with our data---but stars listed as 18th or fainter  in Berkhuijsen et al.\ (1988)
should be viewed with suspicion.  The sort of effect visible in Fig.~\ref{fig:berk} is a classic
symptom of problems with sky subtraction, and we were able to confirm that faint stars near
the nucleus (where the background is high) show the largest problem.

\begin{figure}
\epsscale{0.48}
\plotone{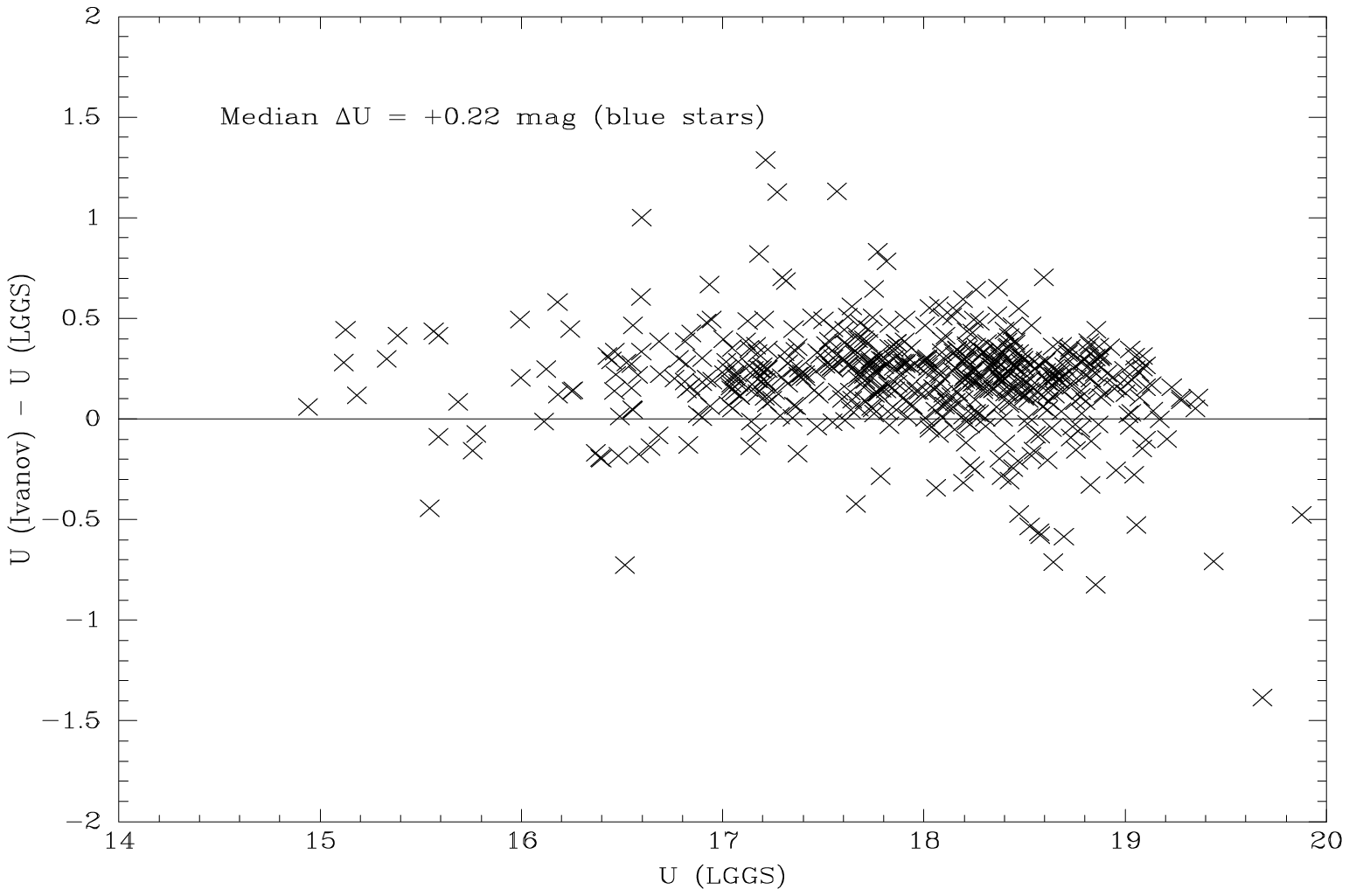}
\plotone{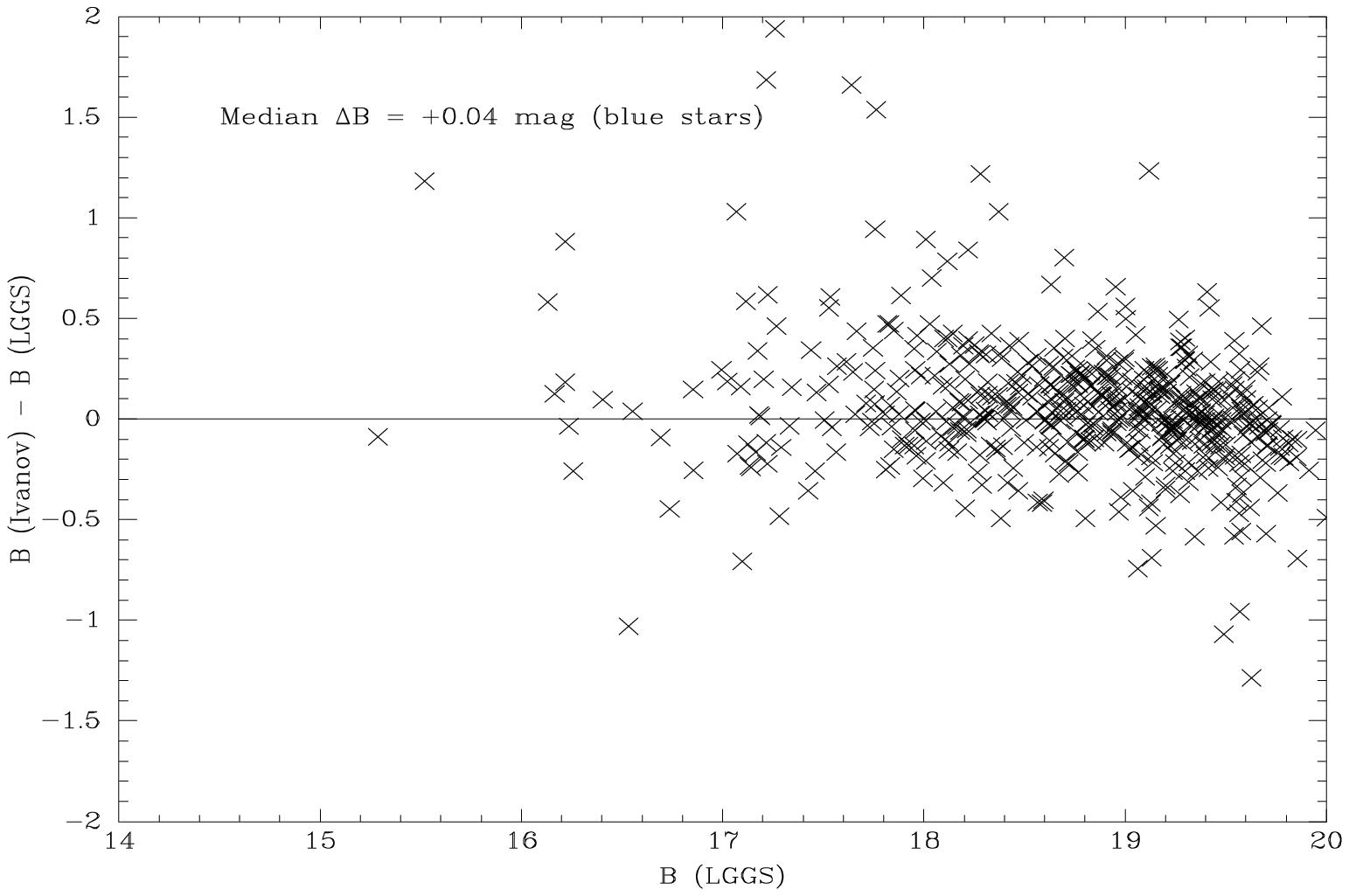}
\plotone{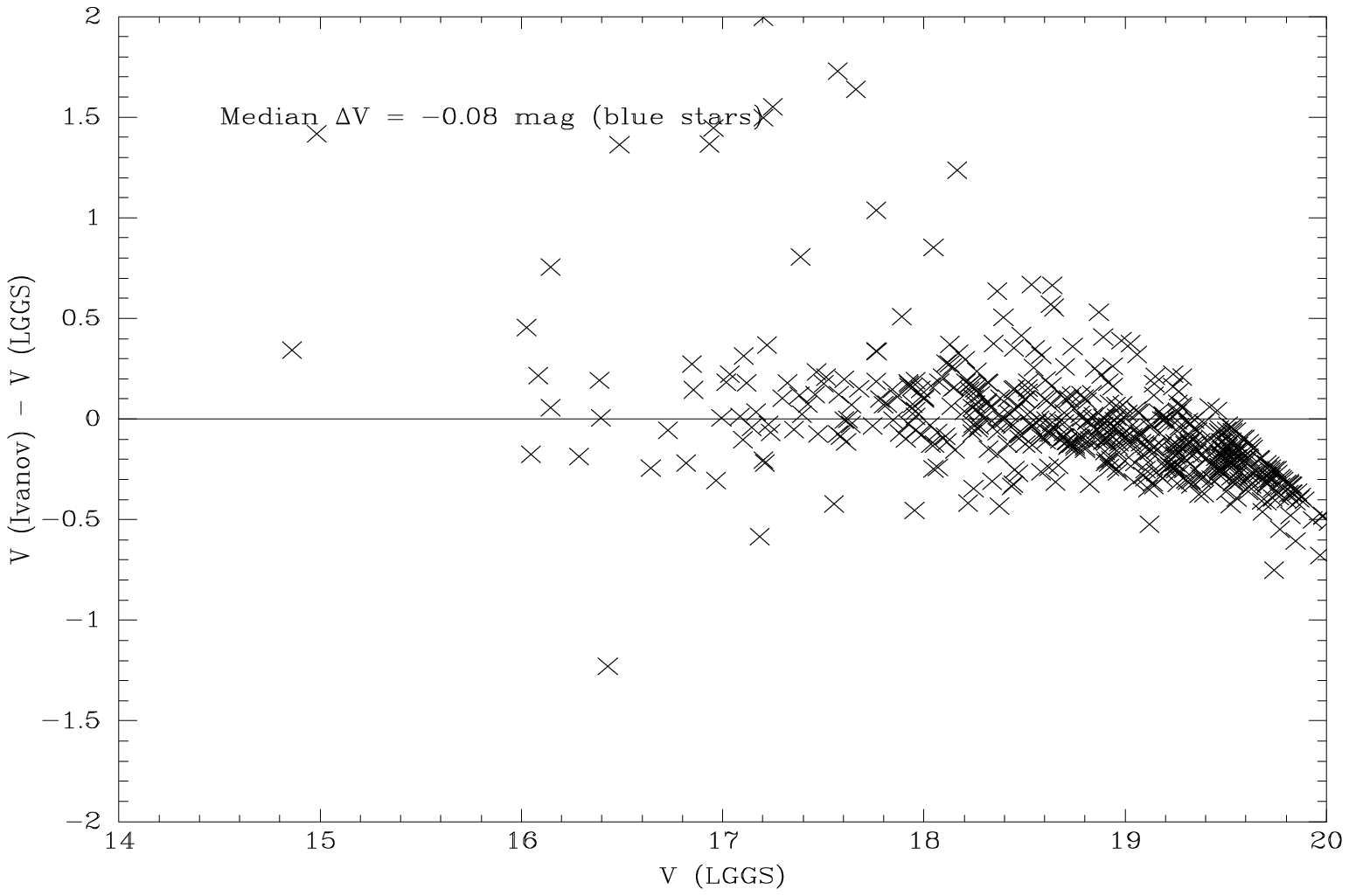}
\plotone{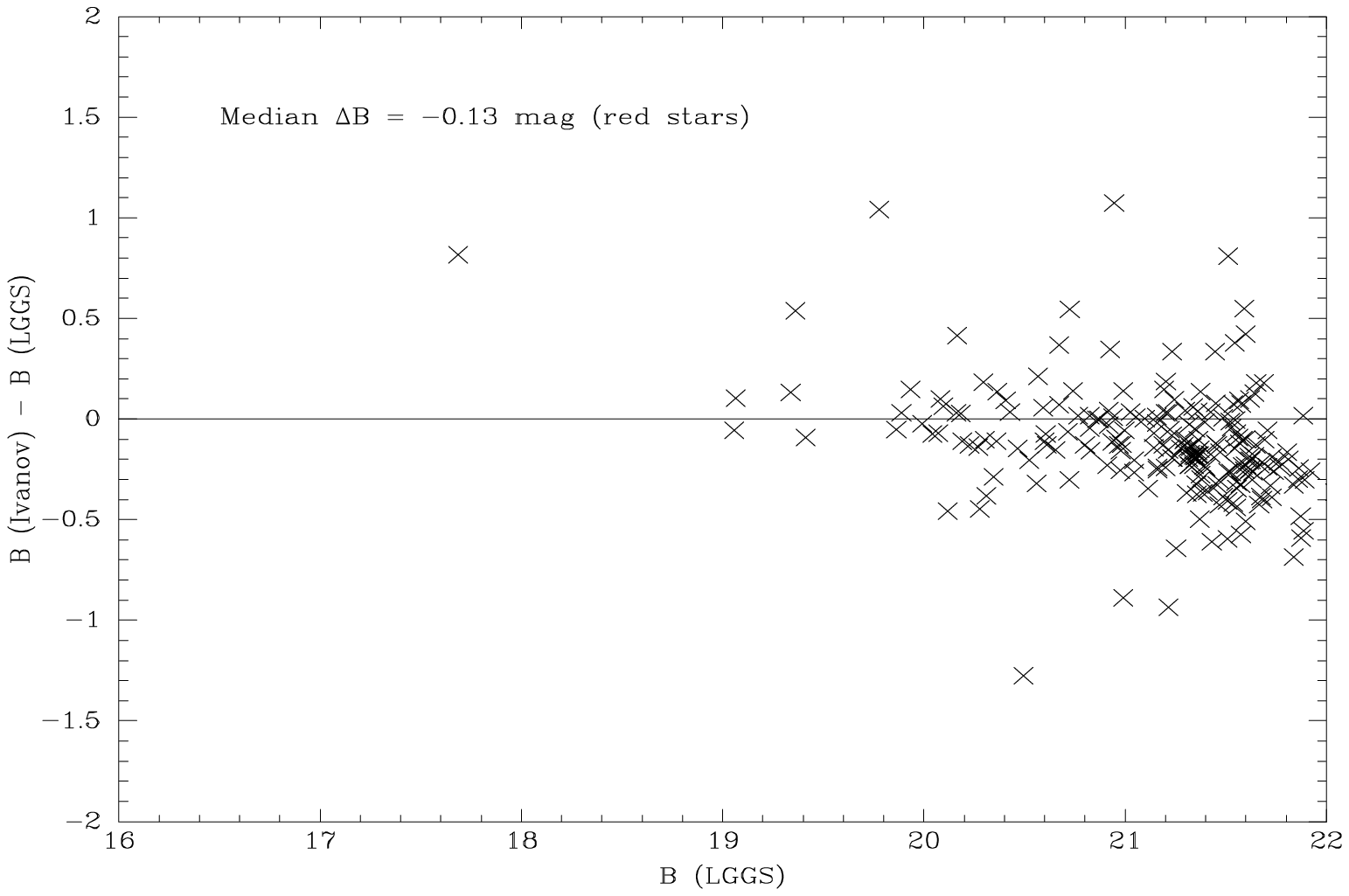}
\plotone{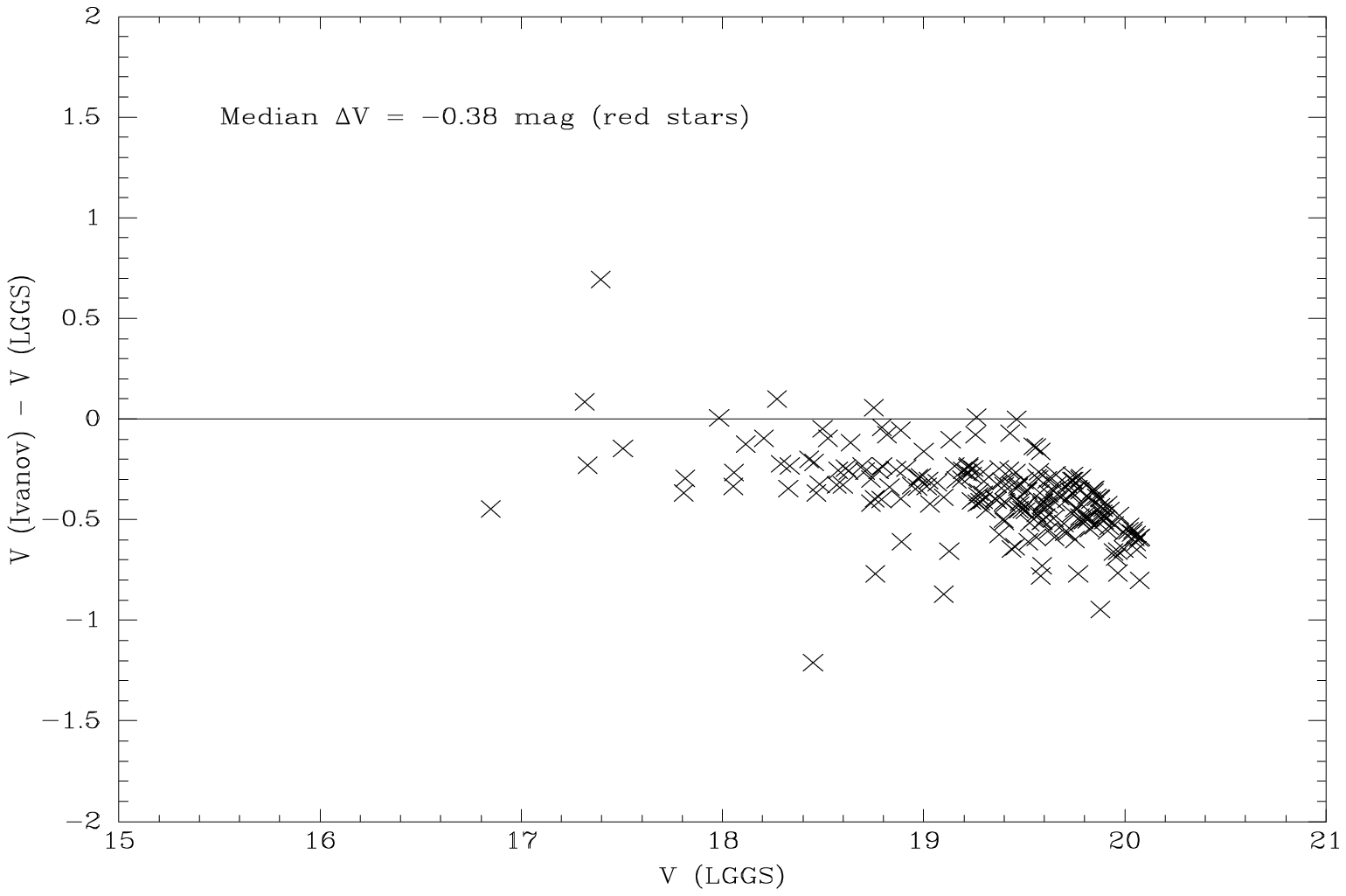}
\caption{\label{fig:ivandif}
Comparison between our photometry and that of Ivanov et al.\ (1993) of M33.
The data for the blue stars show an offset in $U$, and a systematic effect with magnitude for
$V$.  The data for the red stars show systematic problems at the faint end for both $B$ and $V$. }
\end{figure}

The only global set of photometry with coordinates that has been published
 for M33 is that of Ivanov et al.\ (1993),  who present a catalog of blue and red stars, based
upon photographic plates. 
We find
similar problems with those data.  We needed to correct the Ivanov et al.\ (1993)
coordinates by $+0.18^s$ and $-1.1$"; the match against isolated bright stars in our catalog
is usually better than 2.5" after this correction. For the ``blue supergiants" in their
catalog, we find  median differences of $+0.22$ mag in $U$, $+0.04$ mag in $B$,
and $-0.08$ in V, all based upon 558 stars.  As we see in Fig.~\ref{fig:ivandif} the difference in $U$ is
primarily a simple offset, while the differences in $V$ are dominated by the faint stars, which show
a turndown at the faint end ($V>19$).  
This is where we expect errors due to sky determination to be most
severe.  The red stars show a larger effect, with a median difference of $-0.13$ mag in $B$ and
$-0.38$ mag in $V$. (There are no $U$ values for the red supergiants in
Ivanov et al.\ 1993.)

Of course these differences with the photographic studies are not surprising: in their
day they represented the best that could be done, and provided useful photometry and
color information for many years.  The advances brought upon by improved
instrumentation and reduction software result in improved photometry; we hope our
study holds up as well over the next two decades.

\subsection{Spectroscopically Confirmed Members}

We considered providing cross-identification between our stars and those of
others, particularly Magnier et al.\ (1992), who did, after all, provide good coordinates.
While this would be meaningful in the less crowded regions, in the OB associations,
the exact match depends whether a given object is identified as one or more stars.
This is a particular issue given 
 the large difference in seeing between our survey and that of Magnier et al.\ (1992).
 Cross-reference to Berkhuijsen et al.\ (1988) is difficult
 due to the large systematic position
 errors in that catalog, and probably not useful, given the  their
 photometric problems discussed above.
 
Instead, we decided it would be useful to restrict the cross-identifications to stars
spectroscopically confirmed as members of these galaxies.  We present these
in Tables~\ref{tab:M31mem} and \ref{tab:M33mem}, and include the spectral types
and cross-IDs in Tables~\ref{tab:M31} and \ref{tab:M33} as well. 
We began with the spectral types
given in Massey et al.\ (1995, 1996), which includes some earlier work (Humphreys et
al.\ 1990), and updated these with more recently acquired
data by ourselves and others.  Older work, based primarily on photographic spectra, were
added to these (for instance, Humphreys 1979, 1980); since these stars lack published
coordinates, we did the identifications visually, although we restricted this to alleged members,
and ignored the wealth of foreground objects such studies tended to confirm.  We also
included 
``classical" Luminous Blue Variables (LBVs) from Parker (1997).
To this we added recently
 proposed LBV candidates from King et al.\ (1998), Massey et al.\ (1996), and Massey (2006).
 We include the identifications of Wolf-Rayet stars, drawn from Massey \& Johnson (1998).  
 Finally, we include spectroscopically confirmed red supergiants (RSGs), beginning with
 Massey (1998b), and extending back through 
 Humphreys et al.\ (1988).   For the latter, the membership of some stars is 
 questionable.  For instance, Humphrey et al. (1988)'s 
 R79 in M31 is listed as a ``probable supergiant", based upon the strength of the CaII
 triplet. This star is
 also known as OB48-416 (Massey et al.\ 1986); its colors are not those of a
 RSG, which we confirm with our new photometry, but are more like an F-G star, and
 it is likely that the star is a foreground dwarf.  Two additional RSG candidates,
 R138a and R140, fall outside the field covered by our survey.  The identification
 of another RSG candidate, III-R23, was too ambiguous for us to have a positive
 identification.  In general, we concluded that {\it both} radial velocities {\it and} the
 Ca~II triplet strengths were needed to consider a star as a RSG; otherwise, it is
 listed as an ``RSG candidate".  The exceptions were stars with demonstrated
 variability (Var.~66 from van den Bergh et al. 1975, and Vars.\ 4 and 32
 from Hubble 1929.)

 We also indicate in Tables~\ref{tab:M31mem} and \ref{tab:M33mem} whether 
 the object was multiple on our frames.  A star is flagged as ``M" if it has
a companion with $V_{\rm companion}<V_{\rm star}+2.5$ within 1". Multiplicity at this resolution of course
 calls into question the exact identification; which of two stars separated by a
 fraction of an arcsecond dominated an optical ground-based
 spectrum?  In some cases the identifications were uncertain because of poor
 coordinates or finding charts, and we indicate those as well.
 
 In many cases the coordinates are now considerably
 improved (as for the Moffat \& Shara 1983 Wolf-Rayet stars in M31), or, in some 
 cases (such as the spectroscopy of supergiants  in Field IV of
 Baade \& Swope 1963 by Humphreys 1979, or the spectral types of stars in
 M33 from Humphreys 1980) presented for the first time.
 In a few instances we went back to our own original finding charts to ascertain whether
 we had the correct identifications (e.g., M33WR112, M33WR113, M33WR116, and
 M33WR117), 
 which had previously only
 been identified from the poorly reproduced versions of Massey et al.\ (1987a).
 The work also showed that two of the RSGs found by Massey (1998b) in M33 had
 previous spectroscopy by Humphreys (1980).
 Indeed, it was frustration over such identifications that provided some of the
 impetus for the present work.  
 
 Finally, we also
 include in Table~\ref{tab:M31mem} the newly confirmed M31 members
 based upon the spectroscopy presented in \S~\ref{Sec-spectra} and presented
 separately in Table~\ref{tab:M31memnew}.

\section{Analysis}
\label{Sec-analysis}

\subsection{Color-Magnitude Diagrams}
\label{Sec-CMD}

The most fundamental tool at the astronomer's  
disposal for understanding the stellar content of
a region is the color-magnitude diagram (CMD).  In Figs.~\ref{fig:M31CMD} and
\ref{fig:M33CMD} we show the CMDs for M31 and M33, respectively.
We label the regions where we
expect to find the blue and red supergiants, as well as the large central region
where we expect foreground dwarfs and giants to dominate.  The latter is based
upon a consideration of the  Bahcall-Soneira model
(Bahcall \& Soneira 1980) updated by Gary Da Costa and Heather Morrison;
see Massey (2002).  We also
show the confirmed members from
Tables~\ref{tab:M31mem} and \ref{tab:M33mem} with the symbols indicated.
There are of course fewer foreground dwarfs and giants visible against the
face of M33 simply because of the differences in the areas surveyed.  

Several things are apparent from these diagrams.  First, there is clearly a much
more extensive red supergiant population in M33 than in M31.  This effect was
first described by van den Bergh (1973), who noted that the brightest
RSGs in low-metallicity galaxies were brighter relative to the brightest blue
supergiants than in higher metallicity galaxies.  We understand this today as being
primarily due to the effects of mass-loss on the evolution of massive stars; in
high metallicity regions a $30 M_\odot$ star will spend little or none of its He-burning
life as a RSG, but rather will spend it as a Wolf-Rayet star, while in lower metallicity
systems the time spent as a RSG is much longer. (See discussion in Massey 2002,
2003.)  

Secondly, we see that for M31 we expect few if any of the stars identified as 
RSGs by Humphreys (1980), which we have labeled as red supergiant {\it candidates},
to be actual bona fide RSGs.  Instead, they are likely to be foreground objects.
Two of the ``confirmed" RSGs also fall in a peculiar part of the CMD.  The brightest
of these, J004101.4+410434.6 (OB69-46),
has a radial velocity and CaII triplet line strength
consistent with membership, but the $B-V$ color is now 0.4 bluer than the photometry
given by Massey (1998b).  Possibly the identification of this star has been confused.
Four of the RSGs in M33 seem to have a similar problem.

Third, and perhaps the most striking, is that so few of the stars in M31 and M33
have been observed spectroscopically.  The characterization of the stellar populations
of these galaxies has just begun.

\subsection{Identifying the Bluest and Reddest Members from Photometry}
\label{Sec-BR}

\begin{figure}
\epsscale{0.52}
\plotone{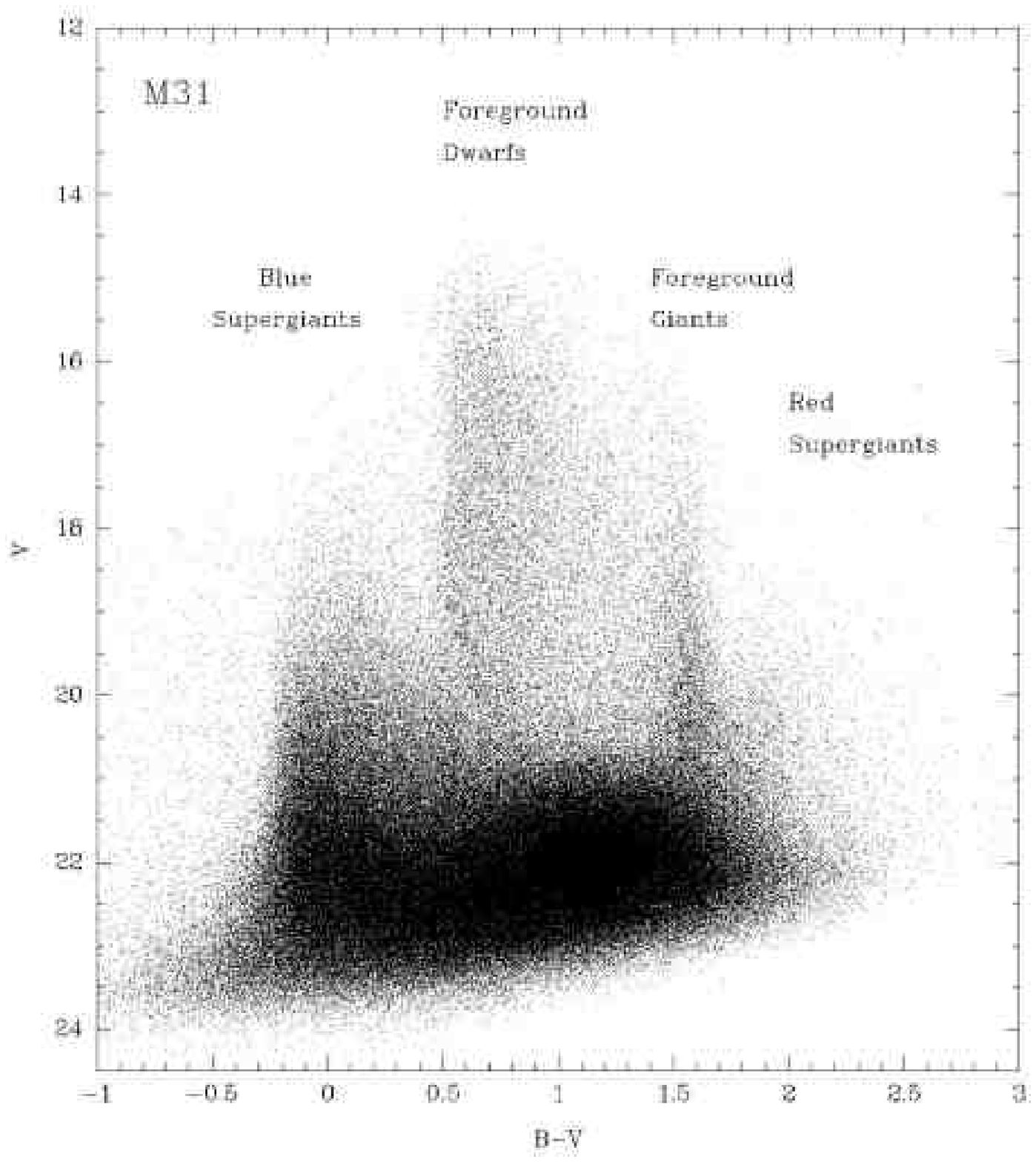}
\plotone{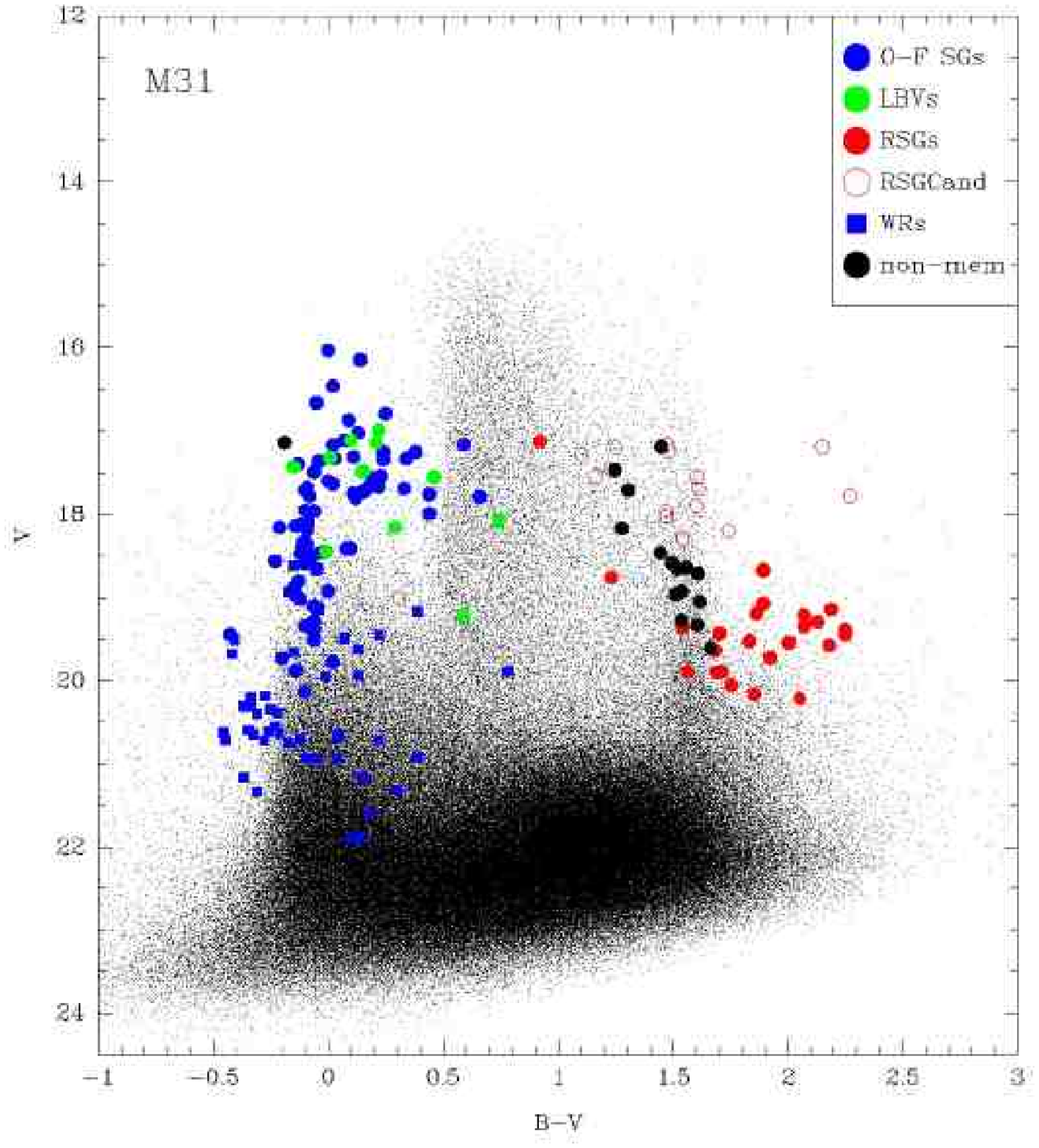}
\caption{\label{fig:M31CMD}
The color-magnitude diagrams for M31.   In the top figure we indicate our
interpretation of the major populations; in the bottom figure, we show the
confirmed members and non-members.  The red open circles (``RSGCand")
denote candidate RSGs, which may or may not be actual members.
} 
\end{figure}

\begin{figure}
\epsscale{0.52}
\plotone{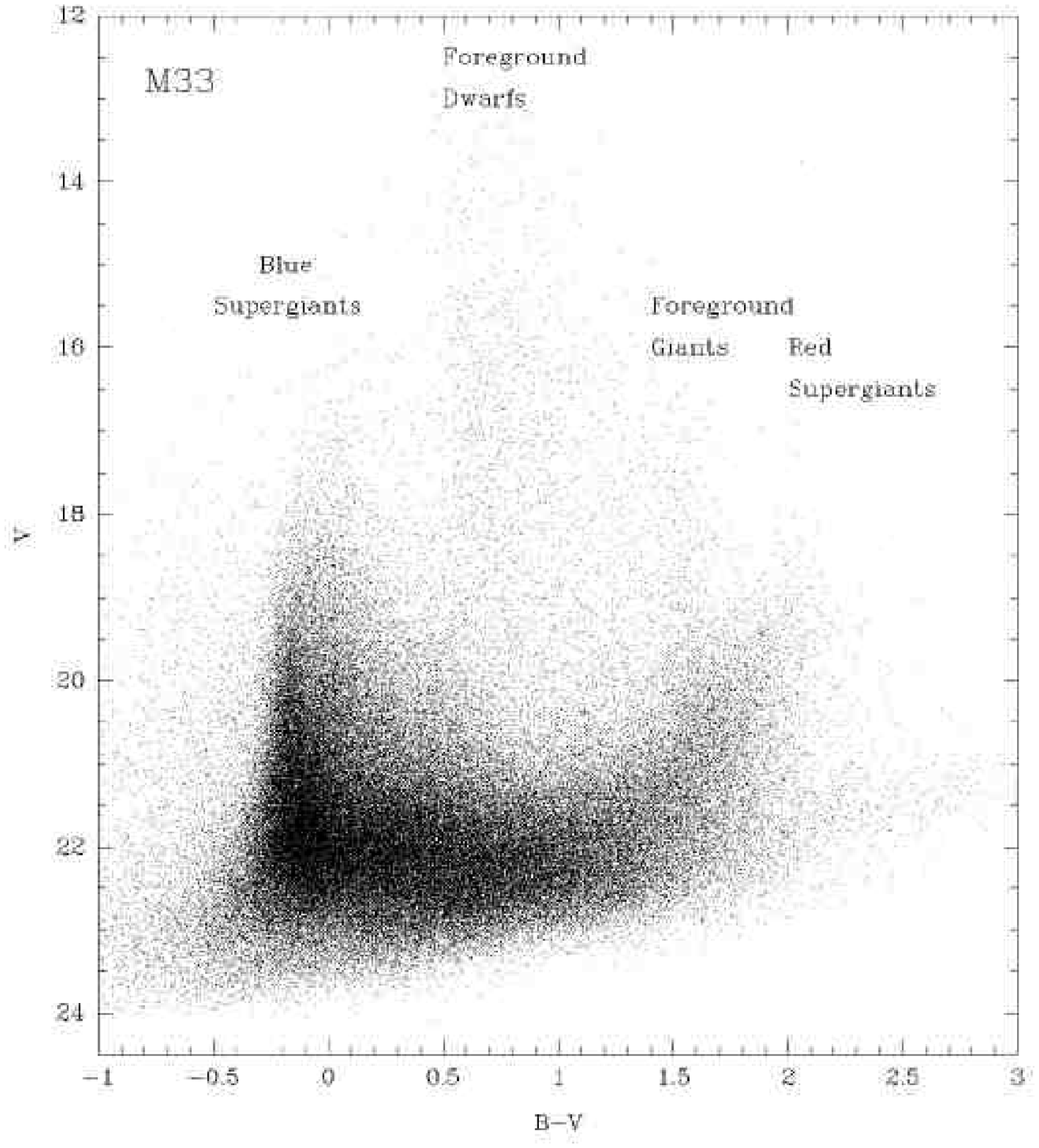}
\plotone{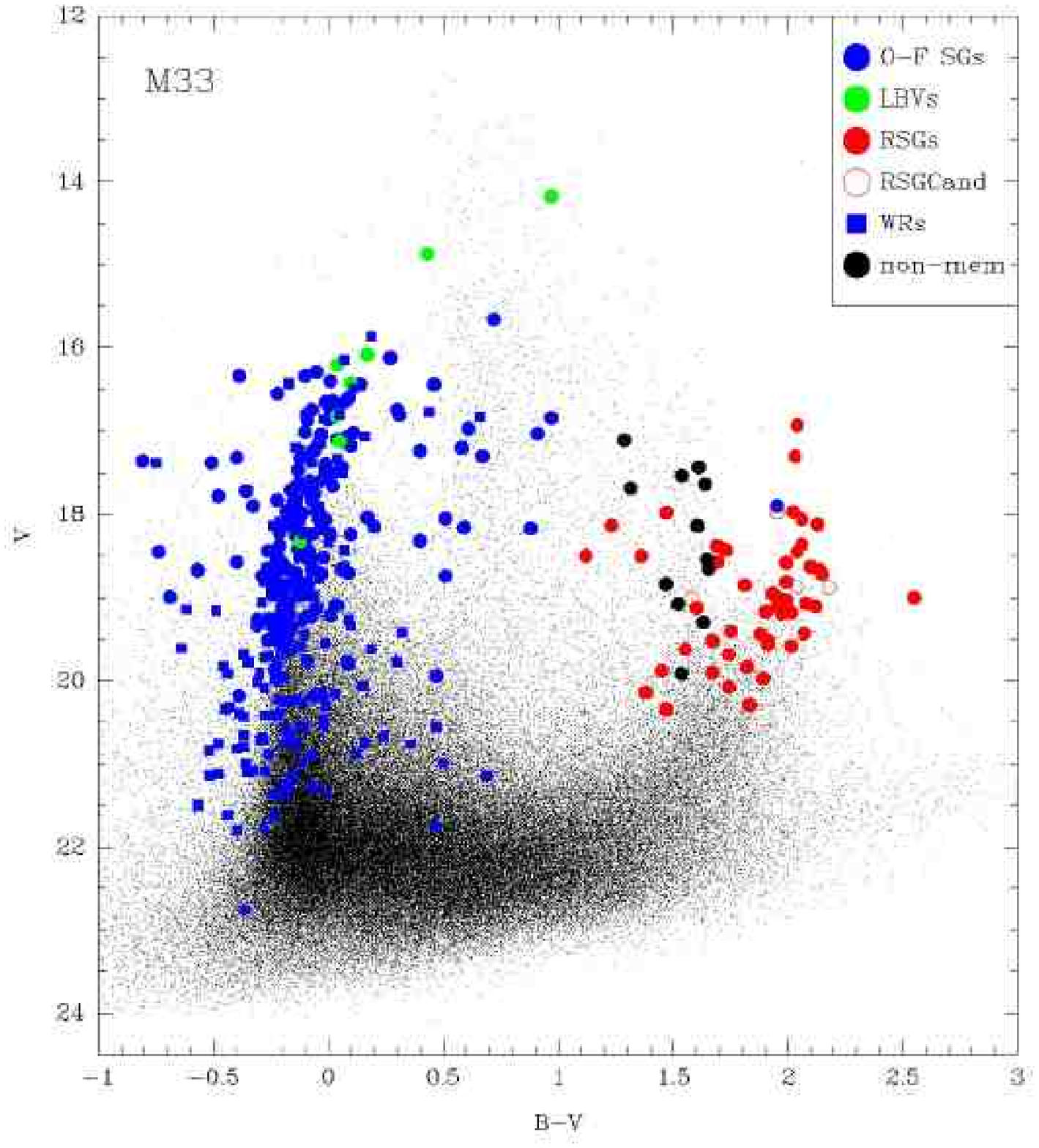}
\caption{\label{fig:M33CMD}
The color-magnitude diagrams for M33.   In the top figure we indicate our
interpretation of the major populations; in the bottom figure, we show the
spectroscopically confirmed members and non-members.
} 
\end{figure}

One of the complications with identifying the hottest massive stars is the issue
of reddening.  In the case of M31 and M33 the foreground reddening is small
and likely uniform ($E(B-V)=0.06$ and 0.07, respectively, according to 
van den Bergh 2000); instead, member stars will be reddened by internal
absorption within the disk of these galaxies.  This adds to some confusion
when trying to separate bona-fide blue supergiants from foreground dwarfs,
a problem apparent in Figs.~\ref{fig:M31CMD} and \ref{fig:M33CMD} where we
find some blue supergiants and LBVs intermixed with the foreground stars.
(Since this class contains some F supergiants, we do expect some overlap.)

The reddening-free Johnson 
{\it Q} index\footnote{$Q=(U-B)-0.72(B-V)$, where we have adopted the canonical
value for the reddening from the Milky Way.} provides a useful
discriminant of intrinsic color, at least for stars with $Q<-0.6$ (earlier than a B2~V or
a B8~I; see Table 3 of Massey 1998c).  For instance, consider a star with $B-V=0.5$
and $V=18$, a region of the CMD that is heavily dominated by foreground dwarfs
(Figs.~\ref{fig:M31CMD} and \ref{fig:M33CMD}).  If $Q=-1.0$ then we can be
assured that the star is a reddened early O-type star, and a member of M31.  If instead
its $Q$ value is $-0.4$, it could be either an unreddened late F foreground dwarf,
or it could be a slightly reddened early A-type supergiant 
member---without spectroscopy
there is no way to tell.

\begin{figure}
\epsscale{0.52}
\plotone{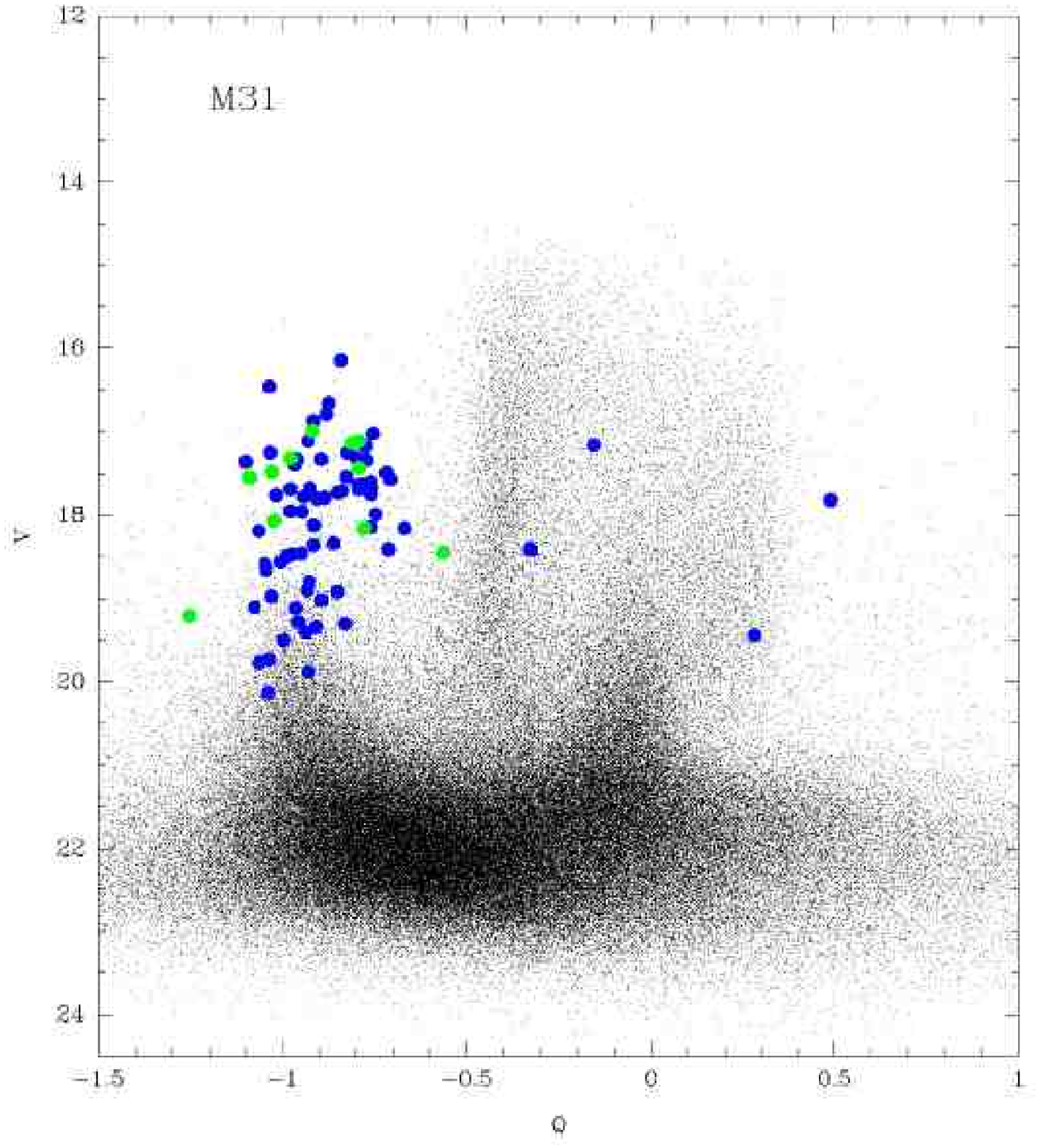}
\plotone{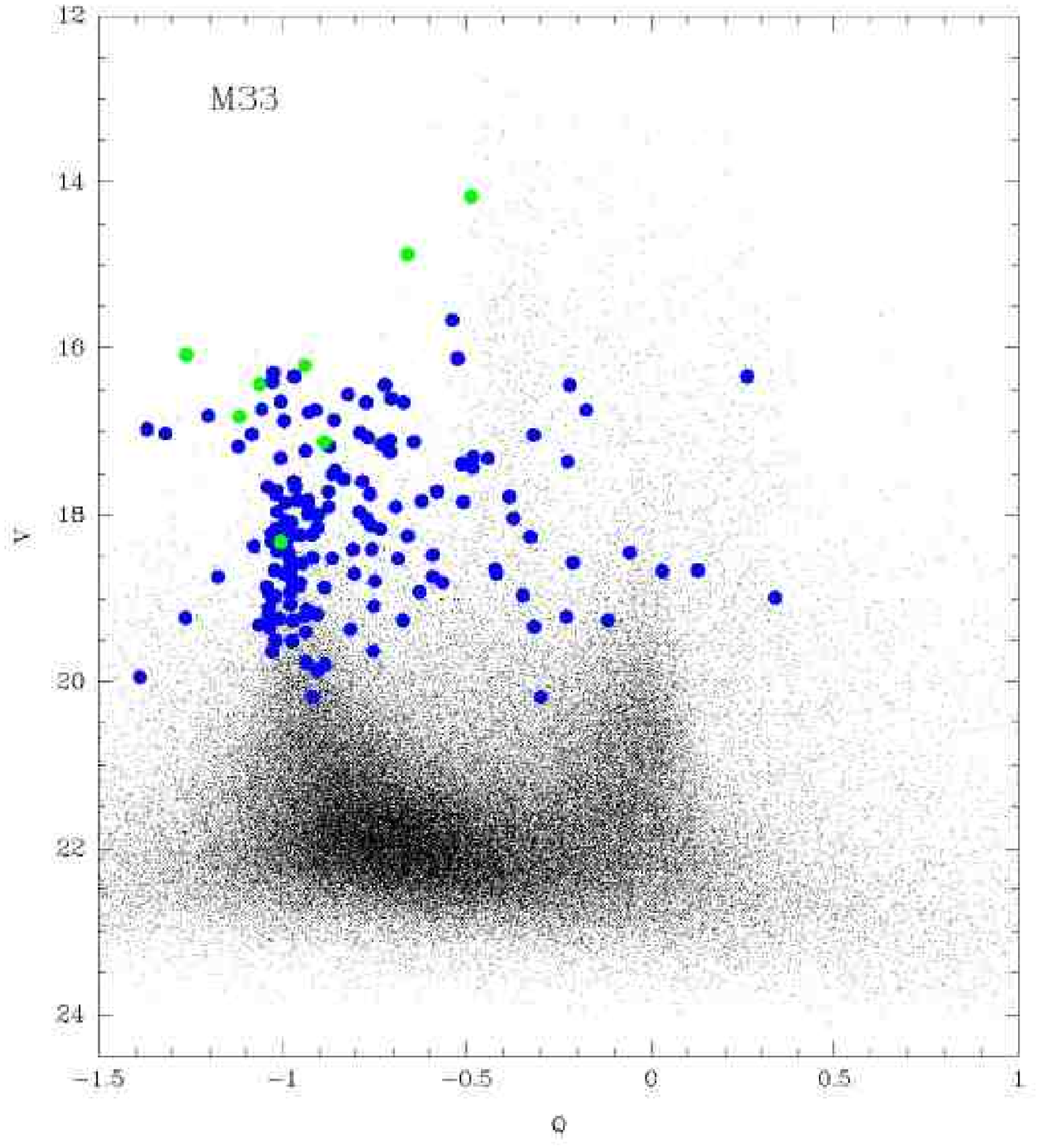}
\caption{\label{fig:QA}
Using $Q$ to find blue supergiants.
The $V$ magnitudes
are plotted against the reddening-free
index $Q$ for M31 and M33.  We indicated the known O-F supergiants (blue 
filled circles) and LBVs (green filled circles).
} 
\end{figure}

In Fig.~\ref{fig:QA} 
we show a $V$ vs $Q$ CMD for each galaxy.  For M31 there is now cleaner
separation between members and non-members, as shown by comparing
Fig.~\ref{fig:QA} with Fig.~\ref{fig:M31CMD}.  
The results for M33 (Fig.~\ref{fig:QA} vs.\ \ref{fig:M33CMD}), which has less internal
reddening (see Massey et al.\ 1995),
 are more ambiguous: the LBVs are now more
obvious, but there is still a significant scattering of blue supergiants into redder
regions of the diagram.  The contrast to M31 is likely due to the fact that simply
a lot more stars have spectroscopy in M33, and that some of these stars are
quite crowded.

\begin{figure}
\epsscale{1.0}
\plotone{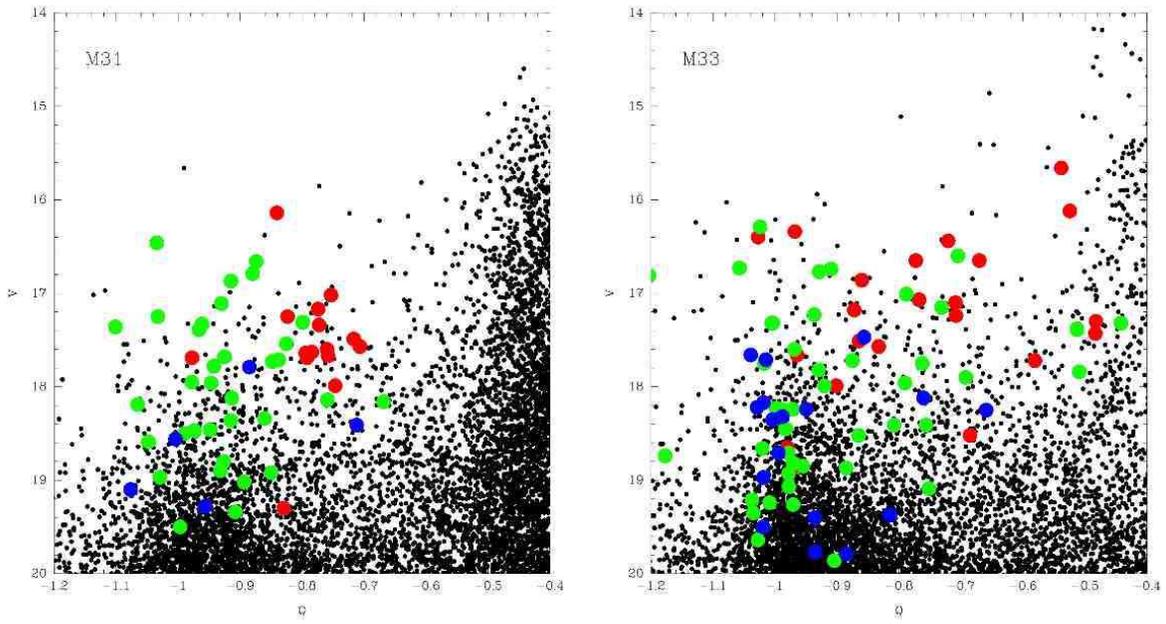}
\caption{\label{fig:QB}
Distinguishing among early-type stars based on $Q$.  The $V$ magnitudes
are plotted against the reddening-free
index $Q$ for M31 and M33.  We indicated the known O-F supergiants
using colored symbols:  O stars (blue filled circles), B0-B3 (green filled circles);
B5 and later (red filled circles).
} 
\end{figure}

However, $Q$ does not prove very useful for distinguishing among the early-type
stars.  In Figs.~\ref{fig:QB} we show an expanded region of the plots,
where we have color-coded (just) the
O-F supergiants by spectral type.  In general, the stars of spectral types B5 and
later can be distinguished from the O's, but in neither galaxy is the separation
clean.  The issue is complicated by crowding (which can affect the photometry
and the derived spectral types) and by the fact that even $Q$ gives only
marginally useful separation (see Massey 1998a).  
The figures illustrate the fact that while the
photometry is good at identifying massive stars (as shown by Figs.~\ref{fig:M31CMD}
and \ref{fig:M33CMD}), quantitative work, such as deriving the IMF, requires
follow-up spectroscopy.

Although distinguishing reddened OB stars from foreground dwarfs is only a minor
problem, it is virtually impossible to identify RSGs on the basis of a single color.
Massey (1998b) found, however, that the two sequences were straight-forward (in
principle) to separate on the basis of a $B-V$ vs $V-R$ two-color diagram.
At a given $V-R$ color, low-surface gravity stars (supergiants) will have a larger
$B-V$ value than will stars with high surface gravities (foreground dwarfs) due to
the effects of line blanketing by weak metal lines in the $B$ bandpass.   This method
should be relatively immune both to reddening and to metallicity; see Fig.~1 of 
Massey (1998b).

\begin{figure}
\epsscale{1.0}
\plotone{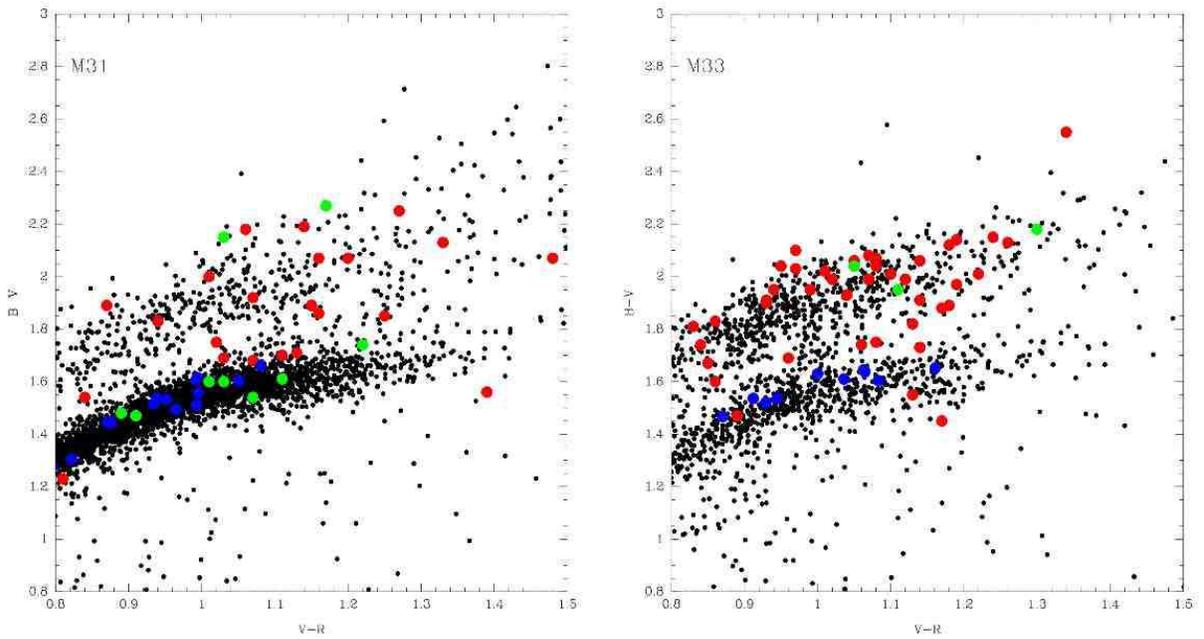}
\caption{\label{fig:rsgs}
Distinguishing red supergiants from foreground dwarfs.  These $B-V$ vs 
$V-R$ plots were made including all stars brighter than $V=20$.  The redder
$B-V$ sequence should correspond to RSGs.  We show red points for
the spectroscopically confirmed RSGs, and  blue points for the stars
with spectroscopy that we have called RSG candidates (Tables~\ref{tab:M31mem}
and \ref{tab:M33mem}),  The green points  are stars that have been
spectroscopically confirmed to be non-members.
} 
\end{figure}

We show such two-color diagrams constructed from our catalogs in 
Figs.~\ref{fig:rsgs}.  First, we see that there is a very clean separation in the
colors, with two easily recognized sequences.  Most of the confirmed RSGs indeed
fall where we expect in this diagram.  A few do not.  It would be worth re-examining
the membership of the outliers.  All of the spectroscopically
confirmed non-members lie where we expect.  

\subsection{Illustrations from a Spectroscopic Reconnaissance}
\label{Sec-spectra}

Characterizing the stellar populations of these two spiral galaxies to the extent
that we can  make useful comparisons with the Magellanic Clouds or the
Milky Way will require a significant amount of new spectroscopy; with 8-m class
telescopes it is now possible to obtain sufficiently high S/N spectra of O
and B stars so that detailed modeling of the physical properties can
be made.  Indeed, such work has already been applied to a few of the brightest
B supergiants in these galaxies (see, for example, Trundle et al.\ 2002). 

However, it is clear from an inspection of Figs.~\ref{fig:M31CMD} and
\ref{fig:M33CMD} that few of the
brightest members have been observed spectroscopically even to the extent
of obtaining of crude spectral types, and establishing membership or non-membership.
The brightest of these can be usefully surveyed on even 4-m class telescopes,
as shown by previous work by Massey et al.\ (1995) and Humphreys et al.\ (1990).

On 29 September 2005 we obtained ``classification" quality spectra of the brightest
M31 stars using the 3.5-m WIYN telescope and Hydra fiber positioner.  The night
was photometric, with good seeing ($\leq$1").  The spectra covered 3970-5030\AA\
in second order, and were obtained with a 790 line mm$^{-1}$ grating (KPC-18C)
with a BG-39 blocking filter with a resolution of 1.5\AA.
 The blue fiber bundle ($\sim 100$ fibers of 3.1" diameter)
was deployed around the $1^\circ$ field of view on targets chosen on the basis
of being blue (reddening-free Johnson $Q<-0.6$) and bright ($V<18$).  
Two fields were observed: a northern one centered at 
$\alpha_{\rm 2000}=00^{\rm h} 44^{\rm m} 20.\!^s6, \delta_{\rm 2000}=+41^\circ37' 00"$
and a southern one centered at
$\alpha_{\rm 2000}=00^{\rm h} 39^{\rm m} 45.\!^s7, \delta_{\rm 2000}=+40^\circ33' 00"$.
The exposure times were 3 hours on each field, in six 30 min exposures.  Halfway through
the sequence the fiber positions were tweaked to take into account changes in the airmass
and hence differential refraction.  The S/N depends upon the star, but typically had a value
of 50 per 1.5\AA\ resolution element.

\begin{figure}
\epsscale{1.0}
\plotone{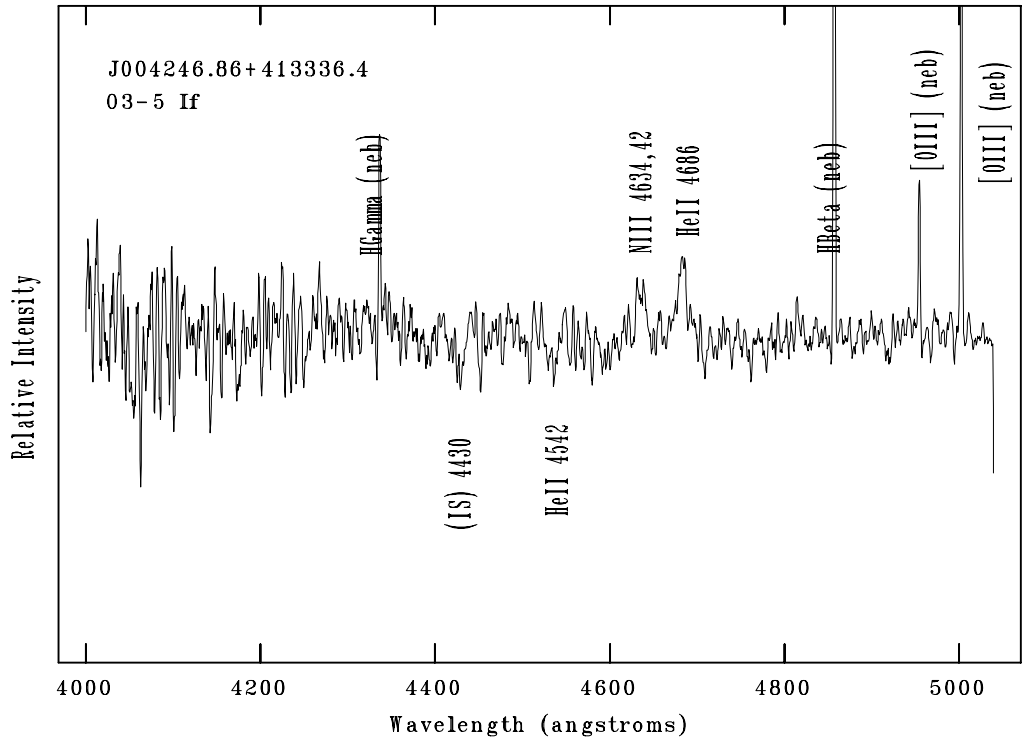}
\caption{\label{fig:M31Ostar}
The O3-5 If star J004246.86+413336.4 in M31.  The lack of He I $\lambda 4471$ relative
to the strong He II $\lambda 4542$ line argues that this star is of early O-type
(O3-5);
the presence of NIII $\lambda 4634,42$ and He~II $\lambda 4686$ leads to the
``If" designation.  This is the earliest O-type star known in M31.
} 
\end{figure}

\begin{figure}
\epsscale{1.0}
\plotone{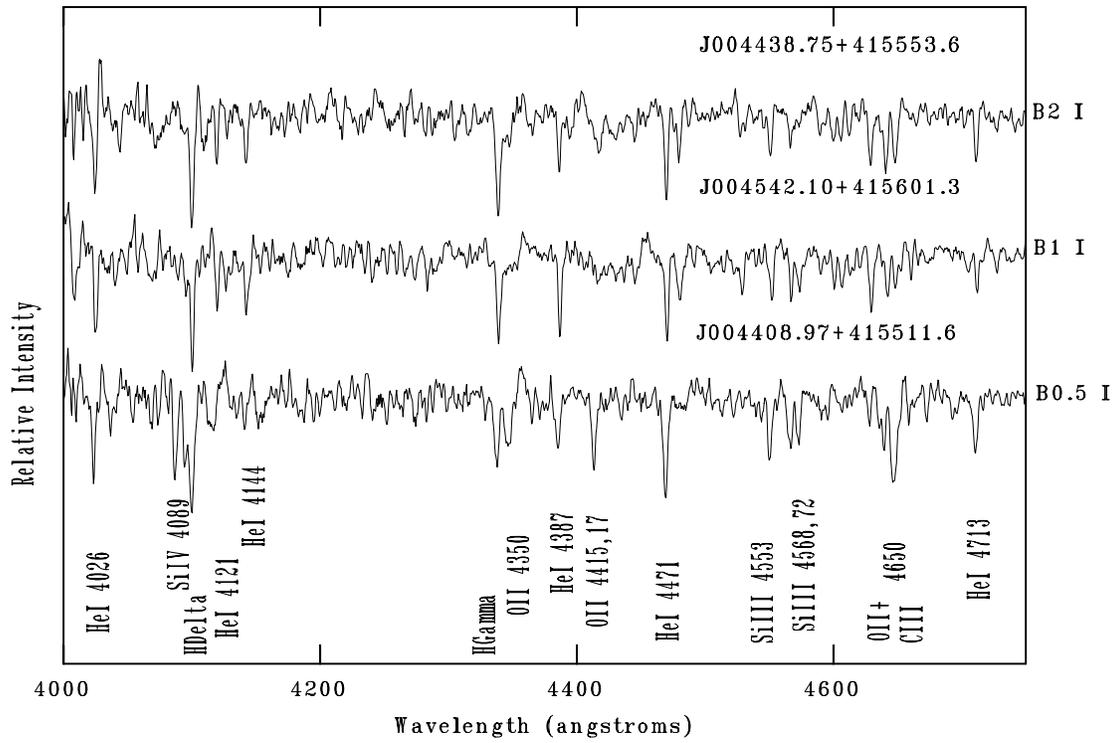}
\caption{\label{fig:M31Bs}
Three early B supergiants in M31.  The principal lines are identified.
} 
\end{figure}

We classified the stars following Walborn \& Fitzpatrick (1990).   We give these
classifications in
Table~\ref{tab:M31memnew}.  As expected, the vast majority
were B supergiants. 
The O stars, although more luminous, are fainter in $V$ because of their
very high effective temperatures and hence significant bolometric corrections.  Nevertheless,
we do find one Of supergiant, as evidenced by the presence of the 
characteristic ``f" emission signature of
NIII $\lambda 4634,42$ and He~II $\lambda 4686$.
(The He~II emission has an equivalent width of -5\AA, well below the -10\AA\ cut-off
usually assigned for Wolf-Rayet stars; see, for example, Massey et al.\ 1987b).
The spectrum lacks the S/N needed for an exact classification, but
given the weakness of He~I, we conclude the star is of spectral type O3-5 If; this
makes it {\it the earliest type O star known in M31.}  We show the spectrum in
Fig.~\ref{fig:M31Ostar}.  We also give some  
representative examples of B supergiants
in Fig.~\ref{fig:M31Bs}.  Only two of the spectra we obtained are foreground dwarfs;
these are also denoted in Tables~\ref{tab:M31} and \ref{tab:M33}.

\begin{figure}
\epsscale{1.0}
\plotone{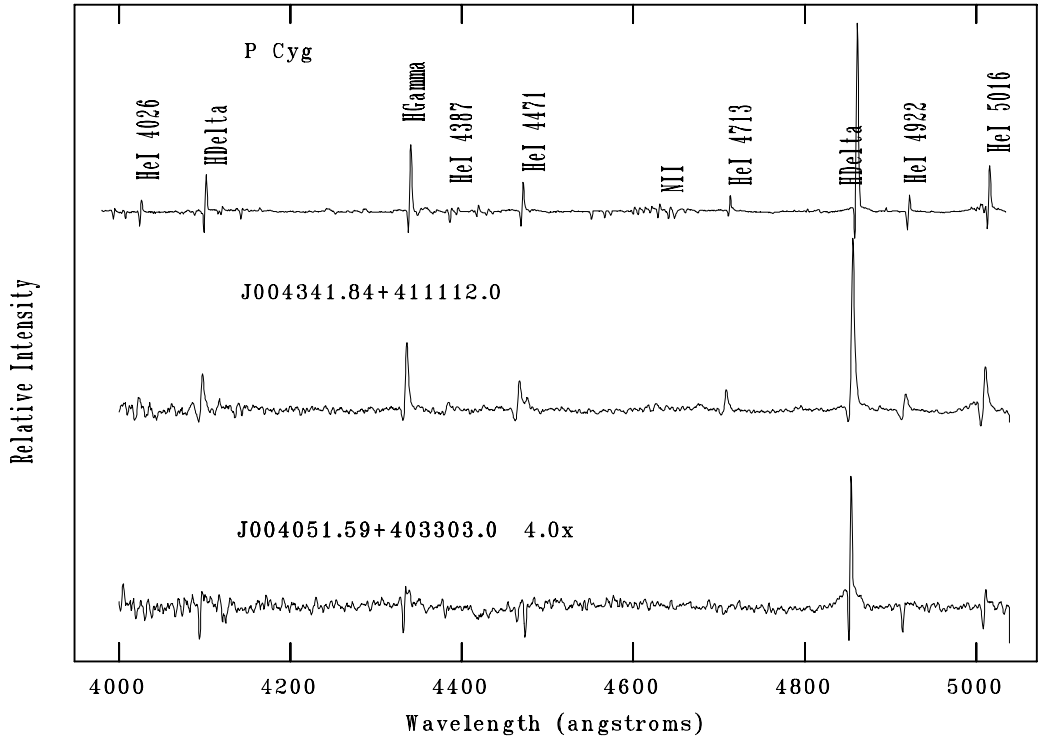}
\caption{\label{fig:PCyg}
Two analogs of P Cygni in M31.  Here we compare the spectra of two newly found
LBV candidates to that of the archetype LBV Galactic star P Cygni.  The scaling
of J004051.59+403303.0 has been enhanced by a factor of 4 to make the features
visible.
} 
\end{figure}

The most interesting discovery is that of two stars with strong P Cygni profiles.
Based upon the spectroscopic similarity  to P Cygni itself, shown in Fig.~\ref{fig:PCyg},
we consider these two stars LBV candidates.  One of these, J004341.84+411112.0,
is the closest known analog to P Cygni, and is discussed in more detail in 
Massey (2006).  Photometrically,
it has been relative constant ($<0.2$~mag) in the optical
over the past 40 years, but with small variations (0.05~mag) seen during a single
year.  Much can be said of the photometric history of P Cygni, which shows only
small variability over the same sort of time-scale; see Israelian \& de Groot
(1999).  An {\it HST} image provides circumstantial evidence of a circumstellar nebula,
bolstering the case (Massey 2006).
The other new LBV candidate, J004051.59+403303.0, is discussed here
for the first time.  The lines are considerably weaker than in P Cygni;
the normalized spectrum has been enhanced by a factor of 4 in Fig.~\ref{fig:PCyg}
to make the lines
visible at the scaling needed for the other two.
Our photometry indicates $V=16.99$, $B-V=0.22$, and $U-B=-0.76$
(all with errors of 0.003~mag)
in 2000.  Magnier et al.\ (1992) observed the star in 1990, and found $V=17.33$,
$B-V=+0.09$, with only slightly larger errors.  Thus, this star seems to be a little
more variable than the first.  The only ``proof" that a star is an LBV is for it to undergo a
dramatic 1-2~mag ``outburst" or show evidence of such a past event in the form
of a circumstellar nebula (see Bohannan 1997); in the meanwhile
we must be content to note the spectroscopic similarity to one of  the archetypes
of LBVs.

\section{Summary and Future Work}
\label{Sec-future}

Our {\it UBVRI} survey of M31 and M33 
produced catalogs containing 371,781 and 146,622 stars.   We achieved our goal of 1-2\% photometry
for the most massive ($>20M_\odot$) stars, with the external photometric calibration providing 
excellent agreement between adjacent fields.  Although the image quality of our data is only
modest (0.8-1.4", median 1.0") by modern standards, our survey covered large areas (2.2 and 0.8 square degrees),
including all of the regions currently known to be actively forming massive stars.  Comparison
of our data with an ACS image of OB 48, a crowded M31 OB association rich in massive stars, suggests
that our catalogs did a respectable job of detecting blends.  

Our color-magnitude diagrams demonstrate the rich stellar content of these systems.  Although
foreground dwarfs and giants will dominate at intermediate colors, most of the stars at either
extreme in color will be blue and red supergiants.  We demonstrate this by providing cross-references
to stars whose spectroscopy has confirmed their memberships in these systems.   New spectroscopy
is presented for bright stars in M31, confirming membership for 34 additional members.  Among these
stars are two newly found LBV candidates, many B-A supergiants, and an O star that is the earliest
type known in that galaxy.

Future work is needed to avail ourselves of these beautiful data.   Only a tiny fraction of these stars
have been observed spectroscopically. Follow-up spectroscopic surveys on larger telescopes will
allow us to determine the initial mass functions for numerous regions of star formation, and help
determine if and how the IMF varies with metallicity and other conditions.  High S/N spectra can be
used to model a range of spectral types, helping to establish how metallicity affects fundamental
stellar properties such as effective temperature.   In addition, our team will continue to analyze
our existing data on other Local Group galaxies currently actively forming stars, and compare those
CMDs to those presented here.  

The premise of the NOAO Survey Program was that data such as those presented here should be
useful to others for their own research.  Towards that end we have made our full catalogs and
images available; in addition, we have carefully documented our reduction techniques, and made
our software also available.

\acknowledgments 
Our interest in characterizing the bright resolved stellar populations of Local Group
galaxies has been whetted by the seminal work of Sidney van den Bergh, Allan Sandage, and Roberta Humphreys, with whom we are grateful for correspondence
and conversations over the years.
The basic IRAF procedures for Mosaic data were written by Frank Valdes, while Lindsey Davis
provided the work that led to the determination of the higher-order astrometric solutions.   The
IRAF reduction process was also improved thanks to thoughtful input by Buell Jannuzi.
Without their efforts the task of reducing Mosaic data would have been prohibitively difficult.
In addition, Taft Armandroff provided much scientific guidance in the implementation of the instrument
on the 4-m.  N. King and A. Saha contributed ideas to the original proposal; in addition,
King helped obtain two nights of observations.  We are grateful to Deidre Hunter for
a critical reading of a draft of this paper, as well as constructive suggestions by
the referee.


\begin{deluxetable}{l c c c c c c c c c c c c c c c c c}
\rotate
\setlength{\tabcolsep}{0.02in}
\tabletypesize{\tiny}
\tablewidth{0pc}
\tablenum{1}
\tablecolumns{18}
\tablecaption{\label{tab:journal}Mosaic Observations}
\tablehead{
\colhead{Field}
&\colhead{$\alpha_{\rm 2000}$}
&\colhead{$\delta_{\rm 2000}$}
&&\multicolumn{2}{c}{$U$\tablenotemark{a}}
& \colhead{}
& \multicolumn{2}{c}{$B$\tablenotemark{b}}
& \colhead{}
& \multicolumn{2}{c}{$V$\tablenotemark{b}}
& \colhead{}
& \multicolumn{2}{c}{$R$\tablenotemark{c}}
& \colhead{}
& \multicolumn{2}{c}{$I$\tablenotemark{d}}  \\ \cline{5-6} \cline{8-9} \cline{11-12} \cline{14-15} \cline{17-18}
\colhead{}  & & & & \colhead{Date} & \colhead{DIQ(")} && \colhead{Date} & \colhead{DIQ(")} && \colhead{Date} & \colhead{DIQ(")} && \colhead{Date} & \colhead{DIQ(")}
&& \colhead{Date} & \colhead{DIQ(")}
}
\startdata
M31-F1 &  00 47 02.4 &+42 18 02 &&2000 Oct 02  &1.2 &&   2000 Oct 02 & 1.4 &&  2000 Oct 02 & 1.0 && 2000 Oct 02& 0.9 && 2000 Oct 02& 0.9\\
M31-F2 &  00 46 06.5 &+42 03 28 &&2000 Oct 03  &1.2   && 2000 Oct 03 & 1.1 && 2000 Oct 03  &1.1 && 2000 Oct 03& 0.9  &&2000 Oct 03& 0.9\\
M31-F3 &   00 45 10.6 &+41 48 54&& 2000 Oct 06  &1.2  &&  2000 Oct 03&  1.2 &&  2000 Oct 06  &1.1 && 2000 Oct 06\tablenotemark{e} &1.0   &&2000 Oct 06 &1.0\\
M31-F4 &  00 44 14.7 &+41 34 20 && 2000 Oct 06  &1.1   && 2000 Oct 06 & 1.4   &&2001 Sep 18  &1.1&& 2001 Sep 18 &1.1&&  2001 Oct 06 & 1.1\\
M31-F5 & 00 43 18.8 &+41 19 46 && 2001 Sep 22 &1.0  &&  2001 Sep 22& 0.9   &&2001 Sep 22 &0.9 &&2001 Sep 22 &0.8&&  2001 Sep 22 &0.8\\
M31-F6 & 00 42 22.9 &+41 05 12 && 2002 Sep 11& 0.9  &&  2002 Sep 11& 0.9  && 2002 Sep 22 &0.9 &&2002 Sep 22 &1.1 &&  2002 Sep 22 &1.2\\
M31-F7 & 00 41 27.0 &+40 50 38&&  2002 Sep 10& 0.9  &&  2002 Sep 10& 0.9 &&  2002 Sep 10 &0.9 && 2002 Sep 10 &0.8 &&  2002 Sep 10&0.8\\
M31-F8 &  00 40 31.1 &+40 36 04&&  2000 Oct 02 & 1.2  &&  2002 Sep 11& 0.9  && 2000 Oct 02  & 1.2 && 2000 Oct 02 &1.3   && 2000 Oct 02 &1.1\\
\enddata
\tablenotetext{a}{Series of 5 dithered 600s exposures unless otherwise noted.}
\tablenotetext{b}{Series of 5 dithered 60s exposures unless otherwise noted.}
\tablenotetext{c}{Series of 5 dithered 50s exposures unless otherwise noted.}
\tablenotetext{d}{Series of 5 dithered 150s exposures unless otherwise noted.}
\tablenotetext{e}{Series of 6 dithered exposures.}
\tablenotetext{f}{Series of 10 dithered exposures.}
\tablenotetext{g}{Series of 6 dithered exposures.}
\tablenotetext{h}{Series of 11 dithered exposures.}
\end{deluxetable}

\begin{deluxetable}{l c c c c c c c c c}
\tablewidth{0pc}
\tablenum{2}
\tablecolumns{9}
\tablecaption{\label{tab:colorterms}Color Terms for the KPNO 4-m Mosaic Camera\tablenotemark{a}}
\tablehead{
\colhead{Color}
&\multicolumn{8}{c}{Chip\tablenotemark{c}} \\ \cline{2-9}
\colhead{Term\tablenotemark{b}}
&\colhead{1}
&\colhead{2}
&\colhead{3}
&\colhead{4}
&\colhead{5}
&\colhead{6}
&\colhead{7}
&\colhead{8}
}
\startdata
$K_{U1}$  & $-.032$  &  $-.036$   &  $-.033$ & $ -.018$    &$-.037$   &$-.031$ &  $-.024$  & $-.023$\\
$K_{U2}$  & $ -.278$ &   $-.260$  &   $-.284$ & $ -.320$   & $-.260$  &$ -.250$ & $ -.269$ & $ -.243$\\
$K_B$      &  $-.145$  & $ -.149$   &  $-.178$ & $ -.148$   & $-.129$  &$ -.159$ & $ -.155$ & $ -.177$\\
$K_V$     &  $+.020$  &  $+.016$  &  $+.008$ &  $+.017$   &$ +.032$ &  $+.021$ & $ +.021$  & $+.011$\\
$K_R$     &  $ -.029$  &  $-.059$   &  $ -.037$ &  $-.027$  &  $-.016$  & $-.043$  & $-.051$  &$ -.037$\\
$K_I$       & $+.042$  &  $+.179$   & $+.084$  & $+.110$   & $+.079$  & $+.062$ & $ +.007$  & $+.026$\\
\enddata
\tablenotetext{a}{Typical uncertainties in the color terms are 0.005.}
\tablenotetext{b}{The color terms are defined as follows:
$$u_{\rm Mosaic} = K_{U1} (U-B)_{\rm std} + C_U, (U-B)_{\rm std}>0$$
$$u_{\rm Mosaic} = K_{U2} (U-B)_{\rm std} + C_U, (U-B)_{\rm std}<0$$
$$b_{\rm Mosaic} = K_{B}   (B-V)_{\rm std} + C_B$$
$$v_{\rm Mosaic} = K_{V}    (B-V)_{\rm std} + C_V$$
$$r_{\rm Mosaic} = K_{R}    (V-R)_{\rm std} + C_R$$
$$i_{\rm Mosaic} = K_{I}   (R-I)_{\rm std} + C_I,$$
\noindent
where the $K$'s are the color terms, and $C$'s are the zero-points.
}
\tablenotetext{c}{Numbered as in the Mosaic manual (Jacoby et al.\ 2000), starting with the 
north-eastern chip
and continuing south along the eastern set of four, and then north along the western four.}

\end{deluxetable}

\begin{deluxetable}{l r r r r r r r r r r r r r r}
\tabletypesize{\scriptsize}
\tablewidth{0pc}
\tablenum{3}
\tablecolumns{15}
\tablecaption{\label{tab:difs}Median Differences of Overlapping Fields}
\tablehead{
\colhead{Fields}
& \multicolumn{2}{c}{$V$}
& &\multicolumn{2}{c}{$B-V$}
& & \multicolumn{2}{c}{$U-B$}
& & \multicolumn{2}{c}{$V-R$} 
&  & \multicolumn {2}{c}{$R-I$} \\ \cline{2-3} \cline{5-6} \cline{8-9} \cline{11-12} \cline{14-15}
\colhead{ }
& \colhead{\#Stars\tablenotemark{a}}
& \colhead{$\Delta$\tablenotemark{b}}
& &\colhead{\#Stars\tablenotemark{a}}
& \colhead{$\Delta$\tablenotemark{b}}
& & \colhead{\#Stars\tablenotemark{a}}
& \colhead{$\Delta$\tablenotemark{b}}
& & \colhead{\#Stars\tablenotemark{a}}
& \colhead{$\Delta$\tablenotemark{b}}
& & \colhead{\#Stars\tablenotemark{a}}
& \colhead{$\Delta$\tablenotemark{b}}   \\
}
\startdata
M31-F1$-$M31-F2   &  857 & +0.005 &&   441 &$-$0.005 &&   381 & $-$0.013 &&  579 & +0.003 && 623 & +0.011 \\
M31-F2$-$M31-F3   & 1276 & +0.004 &&  692 &$-$0.001 &&   545 & +0.004 &&  815 &$-$0.009  &&  904 & +0.015 \\
M31-F3$-$M31-F4   & 1447 & +0.001 &&   772 & +0.001 && 606 & +0.003 &&1124 &  0.000 && 1204 & $-$0.009 \\
M31-F4$-$M31-F5   & 1524 & +0.009 &&  704 & $-$0.009 && 634 & +0.003 &&  928 & +0.005 && 1022 & +0.023 \\
M31-F5$-$M31-F6   & 1394 & $-$0.003&&752 & +0.007 && 713 & +0.005 && 809 & 0.000 && 876 & $-$0.017 \\
M31-F6$-$M31-F7   & 1624 & 0.000      && 817 & +0.003 && 786 & +0.007 && 992 & $-$0.006 && 1146 & $-$0.035\\
M31-F7$-$M31-F8   & 1431 & $-$0.001 &&  785 & +0.006 &&  738 & +0.004 &&  892 &$-$0.003 && 1065 & +0.024 \\
M31-F8$-$M31-F9   & 1414 & $-$0.002 &&  754 & $-$0.006 &&  676 & +0.013 &&  998 &$-$0.004 && 1144 & +0.003 \\
\enddata
\tablenotetext{a}{Number of stars in common with statistical errors $\leq$0.01~mag in each field.}
\tablenotetext{b}{Median differences (mags).}
\end{deluxetable}

\begin{deluxetable}{l r r r r r r r r r r r r r r r r r l l}
\pagestyle{empty}
\rotate
\tabletypesize{\tiny}
\tablewidth{0pc}
\tablenum{4}
\tablecolumns{20}
\tablecaption{\label{tab:M31}M31 Catalog\tablenotemark{a}}
\tablehead{
\colhead{LGGS}
& \colhead{$\alpha_{\rm 2000}$}
& \colhead{$\delta_{\rm 2000}$}
& \colhead{$V$}
& \colhead{$\sigma_V$}
& \colhead{$B-V$}
& \colhead{$\sigma_{B-V}$}
& \colhead{$U-B$}
& \colhead{$\sigma_{U-B}$}
& \colhead{$V-R$}
& \colhead{$\sigma_{V-R}$}
& \colhead{$R-I$}
& \colhead{$\sigma_{R-I}$}
& \colhead{$N_V$}
& \colhead{$N_B$}
& \colhead{$N_U$}
& \colhead{$N_R$}
& \colhead{$N_I$}
&\colhead{Sp.\ Type}
&\colhead{Ref.\tablenotemark{b}}
}
\startdata
J003701.92+401233.2 &00 37  1.92 &+40 12 33.2 &19.862  &0.017& -0.021 & 0.021& -0.928 & 0.015 & 0.204  &0.023 &99.999& 99.999 & 1 & 2&  1 & 1&  0\\
J003701.93+401218.4 &00 37  1.93 &+40 12 18.4 &18.739  &0.008 & 1.494 & 0.015&  0.945 & 0.036  &0.946 & 0.014& 99.999& 99.999 & 1 & 2&  1  &1&  0\\
J003702.03+401141.4 &00 37  2.03 &+40 11 41.4& 21.225 & 0.043 & 1.362 & 0.085& 99.999 &99.999  &0.748  &0.049 & 0.694 & 0.024  &1 & 1&  0&  1&  1\\
J003702.05+400633.5 &00 37  2.05 &+40 06 33.5 &21.091 & 0.044 & 0.050 & 0.061& -1.110 & 0.052  &0.042 & 0.074& 99.999 &99.999 & 1&  2&  1 & 1&  0\\
J003702.13+400945.6 &00 37  2.13 &+40 09 45.6 &16.091 & 0.006 & 1.287&  0.007 & 0.983 & 0.007  &0.792  &0.010 &99.999 &99.999  &1 & 2&  1 & 1&  0\\
J003702.24+401225.7 &00 37  2.24 &+40 12 25.7 &20.765 & 0.029&  1.584&  0.072& 99.999& 99.999  &1.030 & 0.036 & 1.262 & 0.021  &1 & 2&  0 & 1&  1\\
J003702.38+400529.5 &00 37  2.38 &+40 05 29.5 &22.427 & 0.165  &1.359 & 0.498 &99.999 &99.999 & 0.574&  0.180 & 0.892 & 0.072 & 1  &1&  0 & 1&  1\\
J003702.44+400723.2 &00 37  2.44 &+40 07 23.2 &22.461 & 0.173 &-0.153 & 0.202& 99.999 &99.999& -0.018 & 0.252 &99.999& 99.999&1 & 2 & 0&  1&  0\\
\enddata
\tablenotetext{a}{The complete version of this table is in the electronic edition of
the Journal.  The printed edition contains only a sample. 
The full version
can also be found at http://www.lowell.edu/users/massey/lgsurvey.}
\tablenotetext{b}{References for spectral types:
1---Present work;
2---Trundle et al.\ 2002;
3---Humphreys 1979;
4---Massey et al.\ 1995;
5---Massey et al.\ 1986;
6---Humphreys et al.\ 1990;
7---Bianchi et al.\ 1994;
8---Massey, unpublished;
9---Hubble \& Sandage 1953;
10---Humphreys et al.\ 1988;
11---Massey 1998b;
12---Massey \& Johnson 1998 and references therein.}
\end{deluxetable}

\begin{deluxetable}{l r r r r r r r r r r r r r r r r r l l}
\pagestyle{empty}
\rotate
\tabletypesize{\tiny}
\tablewidth{0pc}
\tablenum{5}
\tablecolumns{20}
\tablecaption{\label{tab:M33}M33 Catalog\tablenotemark{a}}
\tablehead{
\colhead{LGGS}
& \colhead{$\alpha_{\rm 2000}$}
& \colhead{$\delta_{\rm 2000}$}
& \colhead{$V$}
& \colhead{$\sigma_V$}
& \colhead{$B-V$}
& \colhead{$\sigma_{B-V}$}
& \colhead{$U-B$}
& \colhead{$\sigma_{U-B}$}
& \colhead{$V-R$}
& \colhead{$\sigma_{V-R}$}
& \colhead{$R-I$}
& \colhead{$\sigma_{R-I}$}
& \colhead{$N_V$}
& \colhead{$N_B$}
& \colhead{$N_U$}
& \colhead{$N_R$}
& \colhead{$N_I$}
&\colhead{Sp.\ Type}
&\colhead{Ref.\tablenotemark{b}}
}

\startdata
J013146.16+301855.6 &01 31 46.16 &+30 18 55.6 &19.555 & 0.013  &1.533  &0.025  &1.141&  0.047 & 1.030 & 0.015& 99.999 &99.999&  1&  1&  1&  1&  0\\
J013146.18+302932.4 &01 31 46.18 &+30 29 32.4 &20.560  &0.068 & 0.645 & 0.117 &99.999& 99.999 & 0.564&  0.115& 99.999 &99.999&  1&  1& 0 & 1&  0\\
J013146.18+302931.4 &01 31 46.18 &+30 29 31.4 &21.027  &0.061 & 0.090 & 0.113  &0.266 & 0.118 & 1.012  &0.111& 99.999& 99.999&  1&  1&  1 & 1 & 0\\
J013146.20+302706.2& 01 31 46.20& +30 27  6.2& 21.057  &0.032 & 1.857 & 0.084 &99.999& 99.999 & 0.924  &0.036 &99.999& 99.999 & 1&  1&  0  &1&  0\\
J013146.21+302026.9& 01 31 46.21 &+30 20 26.9 &21.179  &0.038  &0.962  &0.066 & 0.749&  0.096 & 0.588 & 0.047& 99.999 &99.999  &1&  1&  1&  1&  0\\
J013146.25+301849.7 &01 31 46.25& +30 18 49.7 &16.359 & 0.005  &0.811&  0.007 & 0.470 & 0.007 & 0.460 & 0.007& 99.999& 99.999&  1&  1&  1&  1 & 0\\
J013146.26+302931.5 &01 31 46.26 &+30 29 31.5& 20.247 & 0.063 & 0.912 & 0.128  &0.223&  0.131& -0.209 & 0.112& 99.999& 99.999  &1&  1&  1 & 1&  0\\
J013146.43+302048.4 &01 31 46.43 &+30 20 48.4& 21.990  &0.074 & 0.460 & 0.102 &-0.155&  0.089 & 0.308 & 0.097 &99.999 &99.999&  1&  1&  1&  1 & 0\\
\enddata
\tablenotetext{a}{The complete version of this table is in the electronic edition of
the Journal.  The printed edition contains only a sample.
The full version can also be found at
http://www.lowell.edu/users/massey/lgsurvey.}
\tablenotetext{b}{References for spectral types:
1---Humphreys 1980;
2---Massey et al.\ 1996;
3---Massey et al.\  1995;
4---Massey, unpublished;
5---Monteverde et al.\ 1996;
6---Hubble \& Sandage 1953;
7---van den Bergh et al.\  1975;
8---Massey et al.\ 1998;
9---Massey \& Johnson 1998 and references therein;
10---Catanzaro et al.\ 2003.
}
\end{deluxetable}

\begin{deluxetable}{c c c c c c}
\tablewidth{0pc}
\tablenum{6}
\tablecolumns{6}
\tablecaption{\label{tab:errors}Median Errors\tablenotemark{a}}
\tablehead{
\colhead{Magnitude}
&\colhead{$\sigma_U$}
&\colhead{$\sigma_B$}
&\colhead{$\sigma_V$}
&\colhead{$\sigma_R$}
&\colhead{$\sigma_I$}
}
\startdata
13.0-13.5 & 0.005 & 0.005 & 0.005 & 0.005&  0.001\\
13.5-14.0 &  0.005 & 0.005&  0.005 & 0.005&  0.002\\
14.0-14.5 &  0.005&  0.005& 0.005&  0.004&  0.005\\
14.5-15.0  & 0.005& 0.024& 0.005 & 0.002&  0.005\\
15.0-15.5 &  0.005 & 0.015&  0.005 & 0.002&  0.005\\
15.5-16.0 &  0.005 & 0.010&  0.005&  0.002 & 0.005\\
16.0-16.5 &  0.005 & 0.002&  0.005&  0.002&  0.004\\
16.5-17.0 &  0.005&  0.002&  0.005&  0.002&  0.003\\
17.0-17.5 &  0.004&  0.002&  0.005&  0.002&  0.003\\
17.5-18.0 &  0.003&  0.002&  0.005&  0.002&  0.002\\
18.0-18.5 &  0.004&  0.002&  0.005&  0.003 & 0.002\\
18.5-19.0 &  0.004&  0.003&  0.005&  0.004 & 0.002\\
19.0-19.5 &  0.005&  0.004&  0.005&  0.006 & 0.002\\
19.5-20.0 &  0.006 & 0.006&  0.006&  0.008 & 0.002\\
20.0-20.5 &  0.009&  0.008&  0.009&  0.015&  0.002\\
20.5-21.0 &  0.012&  0.012&  0.014&  0.027&  0.002\\
21.0-21.5 &  0.018 & 0.017& 0.024&  0.038&  0.002\\
21.5-22.0 &  0.027& 0.029&  0.044&  0.048& 0.002\\
22.0-22.5 &  0.042&  0.054&  0.073 & 0.071&  0.013\\
22.5-23.0 &  0.079&  0.100&  0.103 & 0.153&  0.042\\
23.0-23.5 &  0.156&  0.166&  0.255 & 0.285&  0.122\\
23.5-24.0 &  0.285&  0.265&  0.416 & 0.454&  0.234\\
24.0-24.5 &  0.420&  0.397&  0.490 & 0.491&  \nodata\\
24.5-25.0 &  0.495&  0.486 & \nodata &  \nodata&  \nodata\\
\enddata
\tablenotetext{a}{These errors include not only Poisson statistics, but also the
goodness of fit of the PSF.  Thus, the errors of some bright stars can be
higher than that of stars somewhat fainter.}
\end{deluxetable}

\begin{deluxetable}{l c c }
\tablewidth{0pc}
\tablenum{7}
\tablecolumns{3}
\tablecaption{\label{tab:Magnier}Comparison with Magnier et al.\ (1992)}
\tablehead{
\colhead{Property}
&\colhead{LGGS}
&\colhead{Magnier et al.\ (1992)}
}
\startdata
Filters &                   {\it UBVRI}    &  {\it BVRI} \\
Area (square degrees)   &                  2.2          &  1  \\
Seeing(")                &0.8-1.4     &   1.4-4.0  \\
\# Stars\tablenotemark{a}                  &371,781    &  19,966\\
5\% error               & $V=22$    &       $V=20$ \\
PSF-fitting           &DAOPHOT   &      DoPHOT \\
\enddata
\tablenotetext{a}{We include here only stars detected in $B$, $V$, {\it and} $R$.}
\end{deluxetable}

\begin{deluxetable}{l c c r r r r r l c l l}
\pagestyle{empty}
\tabletypesize{\tiny}
\tablewidth{0pc}
\tablenum{8}
\tablecolumns{12}
\tablecaption{\label{tab:M31mem}M31 Members Confirmed by Spectroscopy}
\tablehead{
\colhead{LGGS}
&\colhead{$\alpha_{\rm 2000}$}
&\colhead{$\delta_{\rm 2000}$}
&\colhead{$V$}
&\colhead{$B-V$}
&\colhead{$U-B$}
&\colhead{$V-R$}
&\colhead{$R-I$}
&\colhead{Sp. Type}
&\colhead{Note\tablenotemark{a}}
&\colhead{Cross-ID}
&\colhead{Ref.\tablenotemark{b}}
}
\startdata
&&&& \\                                                                                                                 
\cutinhead{O-F Supergiants}                                                                                             
J003728.99+402007.8&00 37 28.99&+40 20 07.8 &17.31& 0.11&-0.72& 0.06& 0.05&B1I       & &\nodata    &1\\                 
J003733.35+400036.6&00 37 33.35&+40 00 36.6 &18.16&-0.21&-0.82& 0.21& 0.06&B2Ib-B5 I &M&IV-B59     &2\\                 
J003734.36+400116.8&00 37 34.36&+40 01 16.8 &18.59&-0.10&-1.12&-0.01& 0.00&B0 I      & &IV-B24     &3\\                 
J003745.26+395823.6&00 37 45.26&+39 58 23.6 &17.16& 0.59& 0.27& 0.34& 0.40&F5Ia      & &IV-A207    &3\\                 
J003751.90+395901.9&00 37 51.90&+39 59 01.9 &18.41& 0.08&-0.27& 0.07& 0.11&A2 Ib     & &IV-A240    &3\\                 
J004025.48+404423.6&00 40 25.48&+40 44 23.6 &18.47&-0.12&-1.06&-0.02&-0.04&B0.5 Iab  & &OB78-63    &4\\                 
J004028.36+404315.2&00 40 28.36&+40 43 15.2 &18.14&-0.14&-0.86& 0.01&-0.77&B0 Ia     &M&OB78-159   &6\\
\enddata
\tablenotetext{a}{U=Uncertain identification; M=Multiple on a 1" spatial scale.}
\tablenotetext{b}{References for spectral types and cross identifications:
1---Present work;
2---Trundle et al.\ 2002;
3---Humphreys 1979;
4---Massey et al.\ 1995;
5---Massey et al.\ 1986;
6---Humphreys et al.\ 1990;
7---Bianchi et al.\ 1994;
8---Massey, unpublished;
9---Hubble \& Sandage 1953;
10---Humphreys et al.\ 1988;
11---Massey 1998b;
12---Massey \& Johnson 1998 and references therein.
}

\end{deluxetable}

\begin{deluxetable}{l c c r r r r r l c l l}
\pagestyle{empty}
\tabletypesize{\tiny}
\tablewidth{0pc}
\tablenum{9}
\tablecolumns{12}
\tablecaption{\label{tab:M33mem}M33 Members Confirmed by Spectroscopy}
\tablehead{
\colhead{LGGS}
&\colhead{$\alpha_{\rm 2000}$}
&\colhead{$\delta_{\rm 2000}$}
&\colhead{$V$}
&\colhead{$B-V$}
&\colhead{$U-B$}
&\colhead{$V-R$}
&\colhead{$R-I$}
&\colhead{Sp. Type}
&\colhead{Note\tablenotemark{a}}
&\colhead{Cross-ID}
&\colhead{Ref.\tablenotemark{b}}
}
\startdata

\phantom{} \\                                                                                                           
\cutinhead{O-F Supergiants}                                                                                             
J013233.85+302728.9&01 32 33.85&+30 27 28.9 &16.44& 0.46& 0.11& 0.24& 0.26&F8: I     & &R354a      &1\\                 
J013242.92+303847.0&01 32 42.92&+30 38 47.0 &16.73& 0.01&-1.05& 0.02& 0.06&B2.5Ia    & &UIT005     &2\\                 
J013244.97+303457.7&01 32 44.97&+30 34 57.7 &17.01&-0.10&-0.86& 0.12& 0.07&B1.5Ia    &M&UIT007     &2\\                 
J013250.80+303507.6&01 32 50.80&+30 35 07.6 &19.94& 0.47&-1.05& 0.61& 0.79&B1:II:    & &ob21-2     &3\\                 
J013251.74+303527.6&01 32 51.74&+30 35 27.6 &18.95&-0.12&-1.12&-0.02&-0.01&Early B?  & &ob21-8     &3\\                 
J013252.03+303525.1&01 32 52.03&+30 35 25.1 &19.21& 0.01&-1.03& 0.02& 0.08&B0.5Ia    & &ob21-14    &4\\                 
J013252.08+303548.7&01 32 52.08&+30 35 48.7 &19.64&-0.17&-1.15&-0.10&-0.05&B0: I     & &ob21-16    &4\\
\enddata
\tablenotetext{a}{U=Uncertain identification; M=Multiple on a 1" spatial scale.}
\tablenotetext{b}{References for spectral types and cross identifications:
1---Humphreys 1980;
2---Massey et al.\ 1996;
3---Massey et al.\  1995;
4---Massey, unpublished;
5---Monteverde et al.\ 1996;
6---Hubble \& Sandage 1953;
7---van den Bergh et al.\  1975;
8---Massey et al.\ 1998b;
9---Massey \& Johnson 1998 and references therein;
10--Catanzaro et al.\ 2003.
}
\tablenotetext{c}{Coordinates substantially improved for M33WR14, M33WR15, M33WR112, M33WR113, M33WR116, and M33WR117.}
\end{deluxetable}

\begin{deluxetable}{l c c r r r r r l c}
\pagestyle{empty}
\tabletypesize{\tiny}
\tablewidth{0pc}
\tablenum{10}
\tablecolumns{10}
\tablecaption{\label{tab:M31memnew}New Spectral Types M31}
\tablehead{
\colhead{LGGS}
&\colhead{$\alpha_{\rm 2000}$}
&\colhead{$\delta_{\rm 2000}$}
&\colhead{$V$}
&\colhead{$B-V$}
&\colhead{$U-B$}
&\colhead{$V-R$}
&\colhead{$R-I$}
&\colhead{Sp. Type}
&\colhead{Note\tablenotemark{a}}
}
\startdata

J003728.99+402007.8&00 37 28.99&+40 20 07.8 &17.31& 0.11&-0.72& 0.06& 0.05&B1I       & \\
J004028.48+404440.2&00 40 28.48&+40 44 40.2 &17.17& 0.02&-0.76& 0.08& 0.07&B8 I      & \\
J004032.17+404336.7&00 40 32.17&+40 43 36.7 &17.69& 0.17&-0.67& 0.32& 0.70&B5 I      & \\
J004032.37+403859.8&00 40 32.37&+40 38 59.8 &17.76& 0.44&-0.70& 0.31& 0.35&B I       & \\
J004033.80+405717.2&00 40 33.80&+40 57 17.2 &17.33& 0.03&-0.94& 0.10&-0.01&B0.2 I+HII& \\
J004033.90+403047.1&00 40 33.90&+40 30 47.1 &17.75& 0.11&-0.68& 0.11& 0.12&B I       & \\
J004051.59+403303.0&00 40 51.59&+40 33 03.0 &16.99& 0.22&-0.76& 0.22& 0.19&LBVCand!  & \\
J004052.19+403116.6&00 40 52.19&+40 31 16.6 &17.69& 0.33&-0.74& 0.21& 0.23&B8 I      & \\
\enddata
\tablenotetext{a}{F = Foreground star.
}
\end{deluxetable}

\clearpage

\clearpage

\clearpage

\clearpage
\clearpage

\clearpage

\clearpage

\clearpage

\clearpage

\clearpage

\clearpage

\clearpage

\clearpage

\clearpage

\end{document}